\documentclass[11pt,a4paper]{article}
\usepackage{jheppub}
\pdfoutput=1
\usepackage[T1]{fontenc}
\usepackage{bm}
\usepackage{graphicx}
\usepackage{caption}
\usepackage{subcaption}
\usepackage[normalem]{ulem}
\usepackage{gensymb}
\usepackage{slashed}
\usepackage{float}
\usepackage{bbold} 
\usepackage{enumerate}
\usepackage{amsmath}

\usepackage[usenames,dvipsnames]{xcolor}
\RequirePackage[colorlinks=true,urlcolor=blue,anchorcolor=blue,citecolor=blue,filecolor=blue,
              linkcolor=blue,menucolor=blue,linktocpage=true,pdfproducer=medialab]{hyperref}
\definecolor{blus}{cmyk}{1, .9 ,0, 0}
\hypersetup{colorlinks,bookmarksopen,bookmarksnumbered,linkcolor=blus,pdfstartview=FitH,urlcolor=blus,citecolor=blus}
\usepackage[utf8]{inputenc}
\usepackage{multirow}
\usepackage{slashed}

\newcommand{\gr}[1]{{\color{gray} #1}}
\newcommand{\Tr}{{\rm Tr}}

\newcommand{\vev}[1]{\langle #1 \rangle}

\title{Radiative effects in the scalar sector of vector leptoquark models}

\author[a, b]{Rachel Houtz,}    
\emailAdd{rachel.houtz@durham.ac.uk}
\author[c]{Julie Pag\`es}
\emailAdd{julie.pages@physik.uzh.ch}
\author[d]{and Sokratis Trifinopoulos}
\emailAdd{sokratis.trifinopoulos@ts.infn.it}

\affiliation[a]{Department of Physics, Durham University, Durham DH1 3LE, U. K.}

\affiliation[b]{Institute for Particle Physics Phenomenology, Durham University, Durham DH1 3LE, U.K.}

\affiliation[c]{Physik-Institut, Universit\"at Z\"urich, 8057 Z\"urich, Switzerland}

\affiliation[d]{INFN, Sezione di Trieste, SISSA, Via Bonomea 265, 34136, Trieste, Italy}

\date{\today}

\preprint{
IPPP/22/26,
ZU-TH-13/22
}


\abstract{Gauge models with massive vector leptoquarks at the TeV scale provide a successful framework for addressing the $B$-physics anomalies. Among them, the 4321 model has been considered as the low-energy limit of some complete theories of flavor. In this work, we study the renormalization group evolution of this model, laying particular emphasis on the scalar sector. We find that, despite the asymptotic freedom of the gauge couplings, Landau poles can arise at relatively low scales due to the fast running of quartic couplings. Moreover, we discuss the possibility of radiative electroweak symmetry breaking and characterize the fine-tuning 
associated with the hierarchy between the electroweak scale and the additional TeV-scale scalars. Finally, the idea of scalar fields unification is explored, motivated by ultraviolet embeddings of the 4321 model.}

\begin{document}
\maketitle

\phantom{hep-ph/***} 

\vskip 1.5cm


%
%
\section{Introduction}
\label{Sect:Intro}

Recently, there has been a resurgence of interest in the study of leptoquark fields in light of discrepancies in semi-leptonic $B$-decays, collectively known as $B$-physics anomalies~\cite{LHCb:2014vgu,LHCb:2017avl,LHCb:2019hip,LHCb:2021trn,BaBar:2012obs,BaBar:2013mob,Belle:2015qfa,LHCb:2015gmp,LHCb:2017smo,LHCb:2017rln}, which challenge the assumed lepton flavor universal nature of fundamental interactions. 
In particular, the vector leptoquark $U_1$ with Standard Model (SM) quantum numbers $({\bf 3},{\bf 1})_{2/3}$, originally proposed in Ref.~\cite{Alonso:2015sja,Calibbi:2015kma,Barbieri:2015yvd}, has been identified as the only single-mediator solution~\cite{Buttazzo:2017ixm,Angelescu:2021lln}. 
The search for an ultraviolet (UV) completion of the $U_1$ simplified model naturally leads to variations on the Pati-Salam (PS) gauge group, $\mathcal{G}_{\rm PS} \equiv SU(4) \times SU(2)_L \times SU(2)_R$~\cite{Pati:1974yy}, as it features quark-lepton unification. The $U_1$ arises as a massive gauge boson after spontaneous symmetry breaking of the PS group to the SM. However, the original PS model predicts a leptoquark with flavor-universal couplings which has to be very heavy in order to avoid the strict bounds from processes involving the light generations.

The minimal variation containing the $U_1$ leptoquark at TeV scale while being phenomenological viable is $\mathcal{G}_{4321} \equiv SU(4) \times SU(3)' \times SU(2)_L\times U(1)_{X}$ \cite{Diaz:2017lit,DiLuzio:2017vat}, where $SU(2)_L$ is identified with SM weak isospin, and both color and hypercharge are diagonal subgroups (for examples of other variations, see Refs.~\cite{Barbieri:2017tuq,Calibbi:2017qbu,Fornal:2018dqn,Blanke:2018sro}). Models based on $\mathcal{G}_{4321}$ are called \textit{4321 models} and generally fall into two categories based on the charge assignments of the SM matter fields: i) flavor universal 4321~\cite{DiLuzio:2017vat,DiLuzio:2018zxy}, where all would-be SM fermions are singlets under $SU(4)$ and have the usual SM charges and ii) flavor non-universal 4321~\cite{Greljo:2018tuh,Bordone:2018nbg,Cornella:2019hct,Fuentes-Martin:2020bnh,Guadagnoli:2020tlx,Fuentes-Martin:2020hvc,Baker:2021llj}, where the would-be third generation quarks and leptons are unified into $SU(4)$ multiplets. This non-universality in charge assignments translates into non-universality of the couplings of the new heavy vector boson, among them the $U_1$, to the different generations.
The flavor non-universal models can also be viewed as the low-energy limit of more fundamental theories that envision larger symmetry groups broken at various scales and aspiring to dynamically explain the SM Yukawa hierarchies (see e.g. the three-site PS model~\cite{Bordone:2017bld}, PS variations in five dimensions \cite{Fuentes-Martin:2020pww,Fuentes-Martin:2022xnb}
or the twin PS model~\cite{King:2021jeo}). In the following, we will refer to UV models reducing to the 4321 model at low-energy as UV embeddings of 4321.

For both types of 4321 model, the spontaneous symmetry breaking to the SM requires the presence of new scalars at the TeV scale.
The purpose of this work is to investigate the Renormalization Group (RG) evolution of the parameters in the extended scalar sector of the 4321 model. The system of RG equations together with boundary conditions is in general under-determined, since the number of parameters is larger than the number of formal and phenomenological bounds, even in the most simplified version of the model. To study this system, we employ different benchmark scenarios for the boundary conditions. We consider two types of scenarios: i) agnostic towards further assumptions in the UV, and ii) inspired by UV embeddings. The former naturally allows for more freedom in the choice of boundary conditions and is hence suitable for examining the UV behavior in general terms, while the latter introduces additional constraints that have to be satisfied at a certain UV scale. Since we are interested in extrapolating the model as far as possible in the UV, we will focus on the flavor non-universal 4321 model because it has been shown that the universal model exhibits Landau poles at energy scales $\mathcal{O} (100-1000) ~\rm TeV$~\cite{DiLuzio:2018zxy}.

An interesting feature of the setup is the possibility of triggering Electroweak Symmetry Breaking (EWSB) via radiative effects given by the RG flow. Additional scalars coupled to the Higgs sector have the potential to radiatively trigger EWSB, well-established in supersymmetric theories~\cite{Ibanez:1982fr, Inoue:1982pi,Ibanez:1982ee, Ellis:1982wr, Ellis:1983bp, Alvarez-Gaume:1983drc} and examined in non-supersymmetric extensions of the SM in Ref.~\cite{Babu:2016gpg}. In these models, the positive mass squared parameter of the Higgs boson at high energy flows to negative values at low energies, in a similar way as in the Coleman-Weinberg mechanism~\cite{Coleman:1973jx}. Here, we show that the additional TeV scalar states responsible for the 4321 breaking can also radiatively trigger EWSB. We will establish the conditions for radiative EWSB, and in particular, consider the sensitivity of the resulting Higgs mass to the 4321 parameters arranged in the UV.  

The paper is structured as follows. 
In Section~\ref{sec:model} we present the non-universal 4321 model with special emphasis on the scalar potential.
In Section~\ref{sec:analysis} the system of RG equations for the parameters of the scalar sector (Appendix~\ref{app:4321_beta}) is scrutinized. After identifying the most important contributions driving the behavior of the solution in Section~\ref{sec:perturbativity}, we determine the energy scale at which the first Landau pole appears.
 In Section~\ref{sec:REWSB} we identify the conditions required to trigger radiative EWSB and characterize the sensitivity of the physical Higgs mass to the UV values of the relevant parameters.
 Finally, in Section~\ref{sec:UV_conditions} we consider benchmarks satisfying constraints from UV models (Appendix~\ref{app:UV_embedding}) which unify the $\Omega$ fields and the SM Higgs with a second Higgs doublet (Appendix~\ref{app:2HDM}).
We conclude in Section~\ref{sec:conclusions}.

\section{The model}\label{sec:model}

\subsection{Gauge and fermion content}  \label{sec:matter-content}

\paragraph{Gauge structure and heavy vectors.}
The SM gauge group $\mathcal{G}_{\rm SM} = SU(3)_c \times SU(2)_L \times U(1)_Y$, with respective couplings $g_s,g_2,g_Y$, is a subgroup of the 4321 gauge group $\mathcal{G}_{4321} = SU(4) \times SU(3)' \times SU(2)_L \times U(1)_X$, with respective couplings $g_4,g_3,g_2,g_1$. The $SU(4)$ contains $SU(3)_4\times U(1)_4$ as a subgroup. $SU(3)_c$ 
is diagonally embedded in $SU(3)_4 \times SU(3)'$, while $U(1)_Y$ is embedded in $U(1)_4 \times U(1)_X$
with  $Y= \sqrt{\frac{2}{3}}T^{15} + X$ where $T^{15}$ is the rank 4 diagonal generator of $SU(4)$.
The spontaneous symmetry breaking $\mathcal{G}_{4321} \rightarrow \mathcal{G}_{\rm SM}$  gives rise to 15 heavy gauge bosons: a leptoquark $U_1 \sim (\boldsymbol{3},\boldsymbol{1})_{2/3}$, a coloron $G' \sim (\boldsymbol{8},\boldsymbol{1})_0$, and a $Z' \sim (\boldsymbol{1},\boldsymbol{1})_0$, with masses given by
\begin{align}\label{eq:4321_gauge_boson_spectrum}
	\begin{aligned}
		M_{U}&= \frac{g_4}{2} \sqrt{\omega_1^2 + \omega_3^2}\,, \qquad M_{G'}= \sqrt{\frac{g_4^2 + g_3^2}{2}}\,\omega_3 \,,\\
		M_{Z'} &
		= \frac{1}{2\sqrt 6} \sqrt{\left( 3g_4^2 + 2g_1^2 \right)
		\left(
		3 \omega_1^2 + \omega_3^2 \right)
		} \, ,
	\end{aligned}
\end{align}
where $\omega_1$ and $\omega_3$ are the Vacuum Expectation Values (VEVs) of the scalars responsible for the spontaneous symmetry breaking (see Section \ref{sec:scalar_sector}).

Two cases have previously been considered for the VEVs ratio:
\begin{enumerate}[i]
	\item[i)] $\omega_1=\omega_3$ and $g_{1,3}\ll g_4$ as in Refs.~\cite{Fuentes-Martin:2020luw,Fuentes-Martin:2020hvc} in which case there is a residual global symmetry: $SU(4)_V$ custodial from $SU(4)\times SU(4)'$ with $SU(4)'$ the global symmetry containing as subgroup $SU(3)'\times U(1)'$.\footnote{The gauged $U(1)_X$ is then a combination of $U(1)'$ and another $U(1)$, see Ref.~\cite{Fuentes-Martin:2020bnh}.}
	In this case, the heavy vectors have degenerate masses i.e. $M_{U_1}=M_{G'}=M_{Z'}=g_4\,\omega_3 /\sqrt 2$. 
	
	\item[ii)] $\omega_3 > \omega_1$ as in Ref.~\cite{DiLuzio:2018zxy} which is a preferred phenomenological limit
	to avoid significant contributions to
	$\Delta F = 2$ observables.
	In this case and in the limit $g_{1,3} \ll g_4$, we obtain $M_{U_1}= \sqrt{2}M_{Z'}= M_{G'}/\sqrt{2}$. 
\end{enumerate}	
While the limit (i) is particularly motivated in composite scalar sector (see for example Ref.~\cite{Fuentes-Martin:2020bnh}), the limit (ii) offers better compatibility between the case of fundamental scalars and phenomenology. 
For completeness the couplings of the SM gauge bosons, as well as of the heavy bosons $U_1$, $G'$ and $Z'$ are given by~\cite{DiLuzio:2017vat}
\begin{align}
\begin{aligned}
\label{eq:4321_gauge_couplings}
	&g_s =\frac{g_4 g_3}{\sqrt{g_4^2+g_3^2}}~,  &\qquad &g_Y =\frac{g_1g_4}{\sqrt{g_4^2 + \frac{2}{3}g_1^2}}~, &\\
	&g_U =g_4~, &\quad &g_{G'}= \sqrt{g_4^2-g_s^2}~,  &\qquad &g_{Z'}= \frac{1}{2\sqrt 6}\sqrt{g_4^2-\frac{2}{3}g_Y^2} \, .
\end{aligned}
\end{align}

 \paragraph{Fermion content.} 
 In the non-universal 4321 model, only the 3$^{\rm rd}$ generation of quarks and leptons are charged under $SU(4)$. The light generation fields, $q_L^i, u_R^i, d_R^i, \ell_L^i, e_R^i$ with $i=1,2$, are singlet under $SU(4)$ and have the same quantum numbers as in the SM, with hypercharge as their $U(1)_X$ charge and $SU(3)_c$ representation as their $SU(3)'$ ones. The $SU(3)'$ of 4321 can therefore be seen as the color group for light generations. 
 The model also contain one or more vector-like fermions $\chi_{L,R}^j$, where $j$ runs over the number of vector-like fermions. After being integrated out, they generate couplings between the $3^{\rm rd}$ and the light generations of the SM fermions. 
The $3^{\rm rd}$ generation fermions are unified into quadruplet with quarks (partners) and lepton (partners) as follows
 	\begin{align}
 	\begin{aligned}
 		\Psi_L &\sim (\boldsymbol 4,\boldsymbol 1,\boldsymbol 2)_0 \equiv \begin{pmatrix}
 			q_L^3 \\ \ell_L^3
 		\end{pmatrix}  \,, \quad
 		&\Psi_R^+&\sim (\boldsymbol 4,\boldsymbol 1,\boldsymbol 1)_{1/2} \equiv \begin{pmatrix}
 			u_R^3 \\ \nu_R^3
 		\end{pmatrix}\,, \quad
 		\\
 		\chi_{L,R}^j& \sim (\boldsymbol 4,\boldsymbol 1,\boldsymbol 2)_0 \equiv \begin{pmatrix}
 			Q_{L,R}^j \\ L_{L,R}^j \end{pmatrix} \,,
 			\quad
 		&\Psi_R^- &\sim (\boldsymbol 4,\boldsymbol 1,\boldsymbol 1)_{-1/2} \equiv \begin{pmatrix}
 			d_R^3 \\ e_R^3
 		\end{pmatrix} \, .
 	\end{aligned}
 	\end{align}
The vector-like fermions $(Q_{L,R}^j,L_{L,R}^j)$ and the 3$^{\rm rd}$ generation fields $(q^3_L,\ell_L^3)$ in the broken phase of 4321 have the same SM charges as the light generations fields.  In the following, we will consider only one vector-like field $\chi_{L,R}$ ($j=1$).
 
 \subsection{Scalar sector}\label{sec:scalar_sector}

\subsubsection{Scalar content and Lagrangian}\label{sec:scalar_content} 
	The scalar content of the 4321 model used in the analysis is presented in Table~\ref{tab:4321scalars}. The scalar potential is 
\begin{equation} \label{eq:Scalar-potential}
   V =  V_\Omega + V_H + V_{\Omega H} \,,
\end{equation}
where $V_\Omega, V_H, V_{\Omega H}$ are defined below.

	\begin{table}[H]
	    \renewcommand{\arraystretch}{1.2}
		\begin{align*}
 		    \begin{array}{|c|c|c|c|c|}
 		         \hline
	          & SU(4) & SU(3)' & SU(2)_L & U(1)_X \\
			\hline 	
			H 		& 1		    	& 1		& 2	    & 1/2	\\
			\Omega_1	& \overline 4	& 1	    & 1		& -1/2  \\
			\Omega_3	& \overline 4	&  3	& 1		& 1/6   \\
			\hline
			\end{array}
		 \end{align*}
 	\caption{Scalar content of the 4321 model.}
 	\label{tab:4321scalars}
 	\end{table}

The potential for the $\Omega_{1,3}$ scalar fields reads~\cite{DiLuzio:2018zxy}
	\begin{align}
V_\Omega	& =  m_{\Omega_3}^2 \Tr [\Omega_3^\dag \Omega_3]
		+ m_{\Omega_1}^2 \Omega_1^\dag \Omega_1  
		+ \frac{\rho_1}{2} ( \Omega_1^\dag \Omega_1 )^2
		+ \frac{\rho_2}{2}  \Tr [\Omega_3^\dag \Omega_3 ] ^2
		+ \frac{\rho'_2}{2}  \Tr [\Omega_3^\dag \Omega_3 \Omega_3^\dag \Omega_3 ]
					\nonumber \\
	&\ \ \ \ \ 
		+ \rho_3 \Tr [ \Omega_3^\dag \Omega_3 ] \Omega_1^\dag \Omega_1
		+ \rho_4\, \Omega_1^\dag \Omega_3 \Omega_3^\dag \Omega_1
		+ \frac{\rho_5}{3!} \left( [\Omega_3\Omega_3\Omega_3\Omega_1]  + \rm h.c. \right) \, ,
	\label{eq:Omegapotential}
	\end{align}
where $ [\Omega_3\Omega_3\Omega_3\Omega_1] = \epsilon_{\alpha \beta \gamma \delta} \epsilon^{abc} (\Omega_3)^{ \alpha }_{ a } (\Omega_3)^\beta_b (\Omega_3)^\gamma_{ c } (\Omega_1)^{\delta }  $.
Notice that the $\rho_5$ term in Eq.~\eqref{eq:Omegapotential} breaks explicitly a $U(1)$ global symmetry of the scalar potential, thus preventing the appearance of a massless Goldstone boson after spontaneous symmetry breaking.

The Higgs potential is the same as in the SM
\begin{equation} \label{eq:Higgs-potential}
    V_H	 = \mu_H^2 H^\dag H 
		+\frac{\lambda}{2}  ( H^\dag H )^2 \,,
\end{equation}
and the mixing terms between the Higgs sector and the $\Omega$ sector are
\begin{equation} \label{eq:portal-potential}
    V_{\Omega H}	 = 
    \eta_1\,  H^\dag H\,  \Omega_1^\dag \Omega_1
		+ \eta_3\,  H^\dag H\, \Tr [ \Omega_3^\dag \Omega_3 ] \, .
\end{equation}
The roles of each scalar are as follows (the explicit Yukawa terms are written in Appendix~\ref{app:Yukawa}):
	\begin{itemize}
    \item $\Omega_3$ breaks 4321. After getting a vev, since this scalar is charged under both $SU(3)$ light and $SU(4)$ heavy, it will mix the light generation quarks with the vector-like quarks. Therefore, it is responsible for the mass mixing between light quarks and the $3^{\rm rd}$ family quarks, and also for the small coupling between the $U_1$ leptoquark, the colorons and the $Z'$ with the light quarks. 
    \item $\Omega_1$ breaks 4321. After getting a vev, since this scalar is charged under $SU(4)$ heavy and has the right $U(1)_X$ charge, it will mix the light generation leptons with the vector-like leptons. It has the same role as $\Omega_3$ but for leptons instead of quarks (mass mixing between light and $3^{\rm rd}$ family and small coupling of light leptons to heavy gauge boson from 4321 breaking).
    \item $H$ breaks the electroweak symmetry. This scalar is the same as the SM Higgs. It gives mass to all quarks and charged leptons.
    \end{itemize}
The two scalar fields $\Omega_{1,3}$ break spontaneously $\mathcal{G}_{4321} \to \mathcal{G}_{\rm SM}$ by acquiring the VEVs
\begin{equation}
    \langle\Omega_3\rangle = \frac{1}{\sqrt 2}
    \begin{pmatrix}
         \omega_3 & 0 & 0 \\
         0 & \omega_3 & 0 \\
         0 & 0 & \omega_3 \\
         0 & 0 & 0
    \end{pmatrix}
     \quad 
     \text{and}
     \quad 
     \langle \Omega_1 \rangle = \frac{1}{\sqrt 2}\begin{pmatrix}
          0 \\ 0 \\ 0 \\ \omega_1
     \end{pmatrix}\,.
\end{equation}

The $\beta$-functions for the couplings of the potential in Eq.~\eqref{eq:Scalar-potential} are given in Appendix~\ref{app:4321_beta}. 
The scalar content presented in Table~\ref{tab:4321scalars} is the minimal content needed for the analysis. More scalars might be needed for the phenomenology of the 4321 model~\cite{Cornella:2019hct}, however they are not expected to alter significantly the conclusions of this analysis (see Appendix~\ref{app:extended_scalars}).

\subsubsection{Radial and Goldstone modes}

The $\Omega_{1,3}$ scalars can be decomposed as
\begin{equation}
		\Omega_1^\dag =  \begin{pmatrix}
			\tilde T_1 \\
			\frac{\omega_1}{\sqrt 2} + \tilde S_1^*
		\end{pmatrix}\,, \qquad
		\Omega_3^\dag =  \begin{pmatrix}
			\left(\frac{ \omega_3}{\sqrt 2} + \frac{ \tilde S_3^*}{\sqrt 3}    \right)\mathbb{1}_{3} + \,{ \tilde O^{a*}_3} \,t^a \\
			 \tilde T_3^\dag
		\end{pmatrix} \, ,
\label{eq:Omega13}
\end{equation}
where $\mathbb{1}_3$ is the $3 \times 3$ identity matrix and $t^a$ are the $SU(3)$ generators. The complex scalars have the following representations under the SM gauge group: $\tilde S_3 \sim (\boldsymbol 1,\boldsymbol1)_0$, $\tilde T_3\sim(\boldsymbol3,\boldsymbol1)_{2/3}$, $\tilde O_3 \sim(\boldsymbol8,\boldsymbol1)_0$   and $ \tilde S_1 \sim (\boldsymbol1,\boldsymbol1)_0 $, $ \tilde T_1 \sim (\boldsymbol{ 3},\boldsymbol1)_{2/3}$.

The scalar spectrum contains the following mass eigenstates: 
\begin{itemize}
    \item two real octets: the massless $O_{\rm GB}^a=\sqrt{2}\,\text{Im}\,\tilde O_3^a$ and its massive orthogonal combination $O_R^a$;
    \item two complex triplet: the massless $T_{\rm GB} = s_\beta \tilde T_1 -c_\beta \tilde T_3$, with $\tan \beta=\omega_1/\omega_3$, and its massive orthogonal combination $T_R$;
    \item four real singlets, all combination of   $\tilde S_1$ and $\tilde S_3$: the massless $S_{\rm GB}$ and three massive radials $S_0, S_1, S_2$.
\end{itemize}
The Goldstone boson modes $O_{\rm GB}^a$, $S_{\rm GB}$, and $T_{\rm GB}$ fields are eaten by the coloron $G'$, the $Z'$, and the leptoquark $U_1$, respectively and the radial modes acquire the following masses:
\begin{align}
	&\bullet \quad  M_{\, O_R}^2= \omega_3(\rho'_2 \omega_3 - \rho_5 \omega_1) \qquad \qquad   \qquad  
	\text{with} ~  \rho'_2 \omega_3>\rho_5 \omega_1~,  \label{eq:mReO3} \\
	&\bullet \quad   M_{T_R}^2= \frac12 \left(\rho_4  - \rho_5 \frac{\omega_3}{\omega_1} \right) \left(\omega_1^2+\omega_3^2\right) \qquad  
	\text{with}~ \rho_4 \omega_1>\rho_5 \omega_3~, \label{eq:mTR}\\
	&\bullet \quad    M_{S_0}^2= \frac{
	\rho_5}{2} \frac{\omega_3}{\omega_1} \left( 3 \omega_1^2 + \omega_3^2\right) \qquad  \qquad  \qquad  
	\text{with} ~ \rho_5>0~, \label{eq:mS0}\\
    &\bullet \quad   M_{S_1}^2= \frac{1}{2} \left(  \rho_1 \omega_1^2 +  (3 \rho_2 + \rho'_2 )  \omega_3^2 + \frac{\rho_5}{2} \frac{\omega_3}{\omega_1}(\omega_1^2 - \omega_3^2) \right)- \frac{u}{2\omega_1}~,
    \label{eq:mS1}\\
	&\bullet \quad M_{S_2}^2= \frac{1}{2} \left(  \rho_1 \omega_1^2 +  (3 \rho_2 + \rho'_2 )  \omega_3^2 + \frac{\rho_5}{2} \frac{\omega_3}{\omega_1}(\omega_1^2 - \omega_3^2) \right)+ \frac{u}{2\omega_1} \,, \label{eq:mS2}
\end{align}
with $u$ defined as
\begin{align}
	u^2 &= \left[\rho_1\omega_1^3 + (3 \rho_2 + \rho'_2) \omega_1 \omega_3^2 +  \frac{\rho_5}{2} \omega_3 (\omega_1^2 - \omega_3^2)\right]^2 \notag 
 -4 \omega_1 \omega_3 \bigg[  \rho_1\omega_1^3 \left(\frac{\rho_5}{2}  \omega_1 \
+   ( 3 \rho_2 + \rho'_2) \omega_3\right)  \notag \\
&
- 3 \omega_3 \left( \rho_3^2 \omega_1^3 + 
   \rho_3 \rho_5 \omega_3 \omega_1^2 + 
  \frac{\rho_5^2}{3} \omega_3^2 \omega_1+  \frac{\rho_5}{6} (3\rho_2 + \rho'_2) \omega_3^3\right) \bigg]\,. \label{eq:u_par}
\end{align} 

\subsubsection{Electroweak symmetry breaking}
\label{sec:ewsb}

In this section, we identify the conditions for the spontaneous symmetry breaking of the electroweak sector.
Writing the potential Eq.~\eqref{eq:Scalar-potential} in terms of the radial degrees of freedom aligned along the vacua, we obtain
\begin{align} \label{eq:radialspotential}
    \begin{aligned}
    V(s_1, s_3, h) =~ & m_{\Omega_1}^2 s_1^2 +m_{\Omega_3}^2 s_3^2 + \frac{\rho_1}{2} s_1^4 + \frac{1}{6}(3 \rho_2 + \rho_2') s_3^4 + \rho_3 s_1^2 s_3^2 + \frac{2\rho_5}{3 \sqrt 3} s_1 s_3^3 \\
    &+ \frac{\mu_H^2}{2} h^2 + \frac{\lambda}{8} h^4 
    + \frac{\eta_1}{2}h^2 s_1^2 + \frac{\eta_3}{2}h^2 s_3^2 ~ ,
    \end{aligned}
\end{align}
where $s_1 =\frac{\omega_1}{\sqrt 2} \,+ $ Re$(\tilde S_1)$ and $s_3 =\sqrt{\frac32}\, \omega_3 \,+ $ Re$(\tilde S_3)$.

A necessary requirement for our potential to have a minimum at $\vev{h}=v$, $\vev{s_1}= \frac{\omega_1}{\sqrt 2}$  and $\vev{s_3}= \sqrt{\frac{ 3}{ 2}} \omega_3 $
is that the first derivative of the potential evaluated at this point vanishes, i.e. 
\begin{equation}
\left.\frac{\partial V(s_1, s_3, h)}{\partial h_i}\right|_{\vev{h}, \vev{s_1}, \vev{s_3}}=0 \quad\text{for } h_i = h, s_1, s_3 \,.
\end{equation}
These three equations give the set of conditions:
\begin{align}
    \begin{aligned} \label{eq:minimumequation}
    \mu_H^2 + \frac{\eta_1}{2}\omega_1^2 + \frac{3}{2}\eta_3\omega_3^2 + \frac{\lambda}{2} v^2 &=0~,\\
    m_{\Omega_1}^2 + \frac{\eta_1}{2} v^2 + \frac{\rho_1}{2}\omega_1^2 + \frac{3}{2} \rho_3 \omega_3^2 + \frac{\rho_5}{2} \frac{\omega_3^3}{\omega_1} &=0~,\\
    m_{\Omega_3}^2 + \frac{\eta_3}{2} v^2 + \frac{1}{2} (3 \rho_2 + \rho_2') \omega_3^2 + \frac{ \rho_3}{2} \omega_1^2 + \frac{\rho_5}{2} \omega_3 \omega_1 &=0~,
    \end{aligned}
\end{align}
where for each the vanishing VEV solution has been ignored.
From this point forth, since the smallness of $\rho_5$ is technically natural, we neglect its contribution to the effective mass and quartic coupling of the Higgs.  
In the limit where $v \ll \omega_1, \omega_3$, the term proportional to $v^2$ can be omitted in the second and third equations in Eq.~\eqref{eq:minimumequation}.
The Hessian matrix $[M^2_{s_3,s_1,h}]_{ij}=~\dfrac{\partial^2\, V}{\partial h_i \partial h_j}  $ with $h_i \in (s_3,s_1,h)$ evaluated around the stationary points, solutions of Eq.~\eqref{eq:minimumequation}, describes the curvature around these points and give us the conditions for them to be local minima. It is given by
\begin{equation}
\label{eq:massmatrix}
    M^2_{s_3,s_1,h} =
    \begin{pmatrix}
         2 (3 \rho_2 + \rho_2')\omega_3^2  & 2 \sqrt 3 \rho_3 \omega_1 \omega_3& \sqrt 6 \eta_3 v \omega_3 ~,\\
          2 \sqrt 3 \rho_3 \omega_1 \omega_3 & 2\rho_1  \omega_1^2       & \sqrt 2 \eta_1 v \omega_1~,   \\
         \sqrt 6 \eta_3 v \omega_3 & \sqrt 2 \eta_1 v \omega_1  &  \lambda v^2   
    \end{pmatrix} \, .
\end{equation}
The squared mass matrix has to be positive definite, which, for a symmetric matrix, is equivalent to all its leading principal minors being positive
\begin{align} \label{eq:minors}
    \begin{aligned}
        D_1=\,& 3 \rho_2 + \rho_2' >0 \\
        D_{12}=\,& \rho_1 D_1 -3 \rho_3^2 >0 \\
        D_{123}=\,& \lambda D_{12} -  (3\eta_3^2\rho_1  - 6 \eta_1 \eta_3 \rho_3  + \eta_1^2 D_1 )  >0 \, .
    \end{aligned}
\end{align}
The other minors are guaranteed to be positive from the inequalities above, but it is useful to write them explicitly as they put clear constraints on the mixing parameters $\eta_1$ and $\eta_3$, namely
\begin{equation}
      |\eta_1|<\sqrt{\lambda \rho_1} \qquad \text{and} \qquad  |\eta_3|< \sqrt{\frac{\lambda}{3}(3\rho_2 + \rho_2')} \, . 
\end{equation}

In the limit where $v \ll \omega_1, \omega_3$, the eigenvalue of the square mass matrix Eq.~\eqref{eq:massmatrix} corresponding to the physical Higgs mass is
\begin{equation} \label{eq:effectivelambda}
    m_{h_{\rm phys}}^2 = \lambda_{\rm eff} \,v^2 = \frac{D_{123}}{D_{12}}\, v^2 + O \left( \frac{v^2}{\omega_{1,3}^2} \right) \, .
\end{equation}
Given the positive definite conditions Eq.~\eqref{eq:minors}, it is clear from its definition that $\lambda_{\rm eff}>0$.
Moreover, we can define an effective Higgs mass parameter as
\begin{equation} \label{eq:effectivemass}
    \mu_{\rm eff}^2 = -\frac{\lambda_{\rm eff}}{2}v^2 = \mu_H^2 - \frac{\eta_1 D_1 -3 \eta_3 \rho_3}{D_{12}} m_{\Omega_1}^2- 3 \frac{\eta_3 \rho_1 -  \eta_1 \rho_3}{D_{12}} m_{\Omega_3}^2\,.
\end{equation}
The tree-level threshold corrections from integrating out $s_1$ and $s_3$ in the potential Eq.~\eqref{eq:radialspotential} yields an effective potential 
\begin{equation}
    V_{\rm eff} =  \mu_{\rm eff}^2 H^\dag H + \frac{\lambda_{\rm eff}}{2} (H^\dag H)^2~,
\end{equation}
with the same expressions of $\lambda_{\rm eff}$ and $\mu_{\rm eff}^2$ as in Eq.~\eqref{eq:effectivelambda} and Eq.~\eqref{eq:effectivemass}, respectively.
Notice the equivalence between the squared mass matrix diagonalization Eq.~\eqref{eq:massmatrix} with a VEV hierarchy approximation and the tree-level threshold corrections by integrating out the heavy states.\footnote{In Ref.~\cite{Elias-Miro:2012eoi},
the threshold effects in the presence of an additional singlet scalar has been exploited as a mechanism to avoid instabilities of the electroweak vacuum.
} The EWSB happens when $\mu_{\rm eff}^2$ flips sign, from positive to negative.
 
\subsubsection{Bounded from below conditions}

In order for the potential to be Bounded From Below (BFB) and thus for the vacuum to be stable, there must be restrictions on the parameters in the quartic part of the potential $V_4$. 
For the potential $V_\Omega$, the BFB conditions have not previously been derived in the literature, and a full stability analysis of the potential is beyond the scope of this paper. A rigorous treatment of vacuum stability for similar models using the copositivity criteria can be found in Refs.~\cite{Kannike:2012pe,Kannike:2016fmd}. 
Below we list some conditions, obtained by studying the behavior of $V_4$ along specific directions of the fields $\Omega_1$ and $\Omega_3$ and minimizing the potential against extra parameters such as $\theta$ and $\alpha$ as in Ref.~\cite{Chauhan:2019fji}.
 In what follows, we consider strong stability requirements, i.e. $V_4>0$ for arbitrary large values of fields components. Stability in the marginal sense $V_4\geq 0$ (if $V_4= 0$ then one must also require the quadratic part $V_2 \geq 0$) is obtained by making the following inequalities non-strict. For the $\Omega$ potential, we obtained
 \begin{align} \label{eq:Omega_BFB}
    \begin{aligned}
    &|\Omega_3|=0~,&\quad& |\Omega_1| \to \infty  
    & \Rightarrow \quad
    & \rho_1 > 0~, \\
     &|\Omega_3|\to \infty~,&\quad& |\Omega_1|  =0
    & \Rightarrow \quad
    &\rho_2 + \rho_2' > 0~,  \quad 3\rho_2 + \rho_2' > 0~, \\
    &|\Omega_3|^2 = r \cos \theta~, &\quad& |\Omega_1|^2  = r \sin  \theta 
    & \Rightarrow \quad
    &\rho_3  > - \sqrt{\rho_1(\rho_2+\rho_2')}~,\quad  \rho_3 + \rho_4  >- \sqrt{\rho_1(\rho_2+\rho_2')}~,\\
    &\Omega_3 = r e^{i\alpha}~, &\quad& \Omega_1  = r   e^{i\left(\alpha - \frac{\pi}{2}\right)}
    & \Rightarrow \quad
     & |\rho_5| < \frac{1}{4} (\rho_1 + 3 (3\rho_2 +  \rho_2') + 6 \rho_3 )~,
    \end{aligned}
 \end{align}
with $r\to \infty$. For the Higgs potential, the well-known condition is $\lambda >0$.
For the mixed part of the potential $V_{\Omega H}$, using the parametrization $|H|^2= r  \sin \theta$, we obtained
 \begin{align} \label{eq:mix_BFB}
    &|\Omega_3| = 0~, && |\Omega_1|^2  = r  \cos \theta
    & \Rightarrow ~
     & \eta_1 > -\sqrt{\lambda \rho_1}~, \\
    &|\Omega_3|^2 = r  \cos \theta~, && |\Omega_1|  = 0
    & \Rightarrow ~
     & \eta_3 >- \min \left[\sqrt{\lambda (\rho_2 + \rho_2')},\sqrt{\lambda (3\rho_2 + \rho_2')} \right]~. 
     \nonumber
 \end{align}
 
These relations are necessary, but not sufficient, conditions to ensure the vacuum stability.


\section{Analysis of the RG equations system}
\label{sec:analysis}

In this section, we study the RG equations system of the model. In what follows, we explain the procedure for solving this system and list the constraints at the IR boundary. 

\textbf{Methodology.} We solve numerically the system of RG equations using the $\beta$-functions for the 4321 model  presented in Appendix~\ref{app:4321_beta} (neglecting the terms related to the second Higgs doublet) from the scale $\Lambda_{\rm UV}$ to the matching scale $\Lambda_{4}$, and the SM $\beta$-functions from $\Lambda_{4}$ down to $m_Z$. Starting with the full particle content, we first integrate out the heavy gauge bosons of 4321 together with the radials of the scalars $\Omega_{1,3}$ and the vector-like fermion(s) at $\Lambda_{4}$, then the top quark and eventually the Higgs boson at their respective masses. Performing the matching between 4321 and the SM at a common scale is a simplification, because the spectrum of the heavy modes is non-degenerate. However, thanks to the relations imposed by the $SU(4)$ algebra Eq.~\eqref{eq:4321_gauge_boson_spectrum}, the mass differences between the gauge bosons is still expected to be relatively small and no large logarithms are generated. We choose this scale to be the mass of the leptoquark $\Lambda_4 \approx M_U$. Concerning the threshold effects at $\Lambda_4$, we take into account the tree-level corrections due to the decoupling of the heavy scalars $\Omega_{1,3}$ and neglect any subleading loop-level corrections (see Eqs.~\eqref{eq:effectivelambda} and \eqref{eq:effectivemass}).

\medskip

 \textbf{Infrared boundary conditions.}  We require that all studied benchmarks respect the following phenomenological and formal constraints: 
\begin{itemize}
	\item  We initialize the running of the SM gauge couplings, the parameters of the Higgs potential and the top Yukawa coupling as follows~\cite{ParticleDataGroup:2020ssz}:
    \begin{align*}
    \alpha_Y(m_Z)&=0.01017~, \quad \alpha_2(m_Z)=0.03380~, \quad \alpha_s(m_Z)=0.1179~, \\
    \mu_H^2(m_{h_{\rm phys}})&=-(m_{h_{\rm phys}})^2/2~, \quad \lambda(m_{h_{\rm phys}})=0.25923~, \quad y_t(m_t)=0.93419~.
    \end{align*}
    where $\alpha_i = g_i^2 / 4\pi$,  $m_{h_{\rm phys}} = 125.25~\rm GeV$ and $m_t= y_t\, v/\sqrt{2}$ with $v=246$ GeV.
	\item In order to address the $B$-physics anomalies within this framework, assuming both left-handed and right-handed couplings of the leptoquark, the key condition is~\cite{Cornella:2021sby}\footnote{The fit of Ref.~\cite{Cornella:2021sby} includes the charged-current anomalies, encoded by the ratios $R_D^{(*)}$~\cite{BaBar:2012obs,BaBar:2013mob,Belle:2015qfa,LHCb:2015gmp,LHCb:2017smo,LHCb:2017rln} (see Table 2.1), the neutral-current anomalies $R_K^{(*)}$~\cite{LHCb:2014vgu,LHCb:2017avl,LHCb:2019hip,LHCb:2021trn} as well as $B_s \to \mu^+\mu^-$ (see Table 2.2) and a plethora of low-energy constraints (see Table 3.1).}
    \begin{equation}
    \label{eq:gUmU}
    g_U  = (1.1 \pm 0.2) \times\left(  \frac{M_U}{2~{\rm TeV}} \right)~.
    \end{equation} 
    Using Eq.~\eqref{eq:4321_gauge_boson_spectrum} and Eq.~\eqref{eq:4321_gauge_couplings}, the above relation is equivalent to $\sqrt{\omega_1^2 + \omega_3^2} \in [3,4.5] ~\rm TeV$. 
	\item The lower bounds for the masses of the 4321 gauge bosons are set by LHC searches at high $p_T$~\cite{Baker:2019sli,Cornella:2021sby}. Since the masses of the gauge bosons are correlated, the exclusion limits extracted from searches of resonant coloron production with $t\bar{t}$ and $b\bar{b}$ final states impose the leading constraints on the whole mass spectrum: $M_{G'}\gtrsim 4~\rm TeV$ and $M_U,M_{Z'}\gtrsim 3~\rm TeV$ .
	\item The colored radial modes $O_R$ and $T_R$ can be produced via QCD interactions at colliders. In order to avoid direct detection bounds at the LHC, their masses should be above $2~\rm TeV$ (see Ref.~\cite{Diaz:2017lit} for the color triplet and Ref.~\cite{Bai:2018jsr} for the octet). For the singlet radial modes $S_{0,1,2}$, the collider bounds are much milder since the only available ones originate from direct Higgs searches~\cite{Ilnicka:2018def}. Taking the minimum lower mass bound to be $500~\rm GeV$ and the mixing couplings $\eta_{1,3}$ to be $\mathcal{O}(0.1)$, they can be safely avoided. Note that the above mass bounds translate into constraints for the coupling values $\rho_i(\Lambda_4)$ and the VEVs $\omega_{1,3}$ (see Eqs.~(\ref{eq:mReO3}-\ref{eq:mS2})). 
	\item Flavor constraints involving the radial $T_R$, e.g. one-loop FCNC via the Yukawa interaction, impose an important lower bound on the ratio $\tan\beta_T = \omega_3 / \omega_1$. As discussed in Ref.~\cite{DiLuzio:2018zxy}, in case of the universal 4321, the radial $T_R$ together with the vector-like leptons enter box diagrams that generate contributions to the amplitudes of $B_s-\bar{B_s}$ and $D-\bar{D}$ mixing. These contributions are proportional to $\tan \beta_T ^{-4}$ and for $\tan \beta_T \gtrsim 2$ they become negligible. In the present case of the non-universal 4321 only the $D-\bar{D}$ mixing is affected, and still requires a similar suppression realized by assuming a hierarchy of VEVs i.e. $\omega_3 > \omega_1$.
	 \item We require that the BFB conditions Eq.~\eqref{eq:Omega_BFB} and Eq.~\eqref{eq:mix_BFB} are satisfied at all scales.
	 \item The mass parameters $m_{\Omega_{1,3}}^2$ at scale $\Lambda_4$  are set to the minimum of the potential by satisfying Eq.~\eqref{eq:minimumequation}. 
\end{itemize}

\subsection{General dependencies and perturbativity range}
\label{sec:perturbativity}

\begin{figure}
  \centering
  \begin{subfigure}{0.49\textwidth}
  \includegraphics[width=\textwidth]{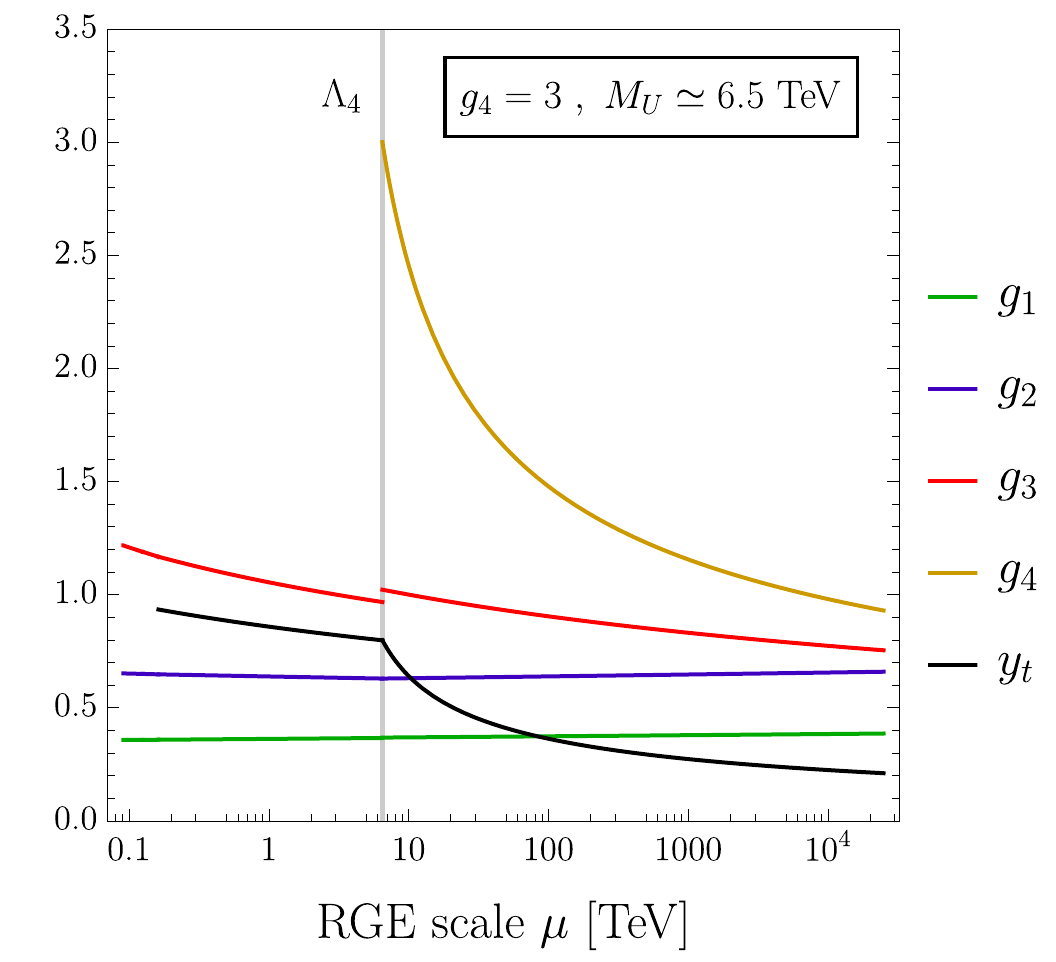}   
    \end{subfigure}
    \begin{subfigure}{0.49\textwidth}
	\includegraphics[width=\textwidth]{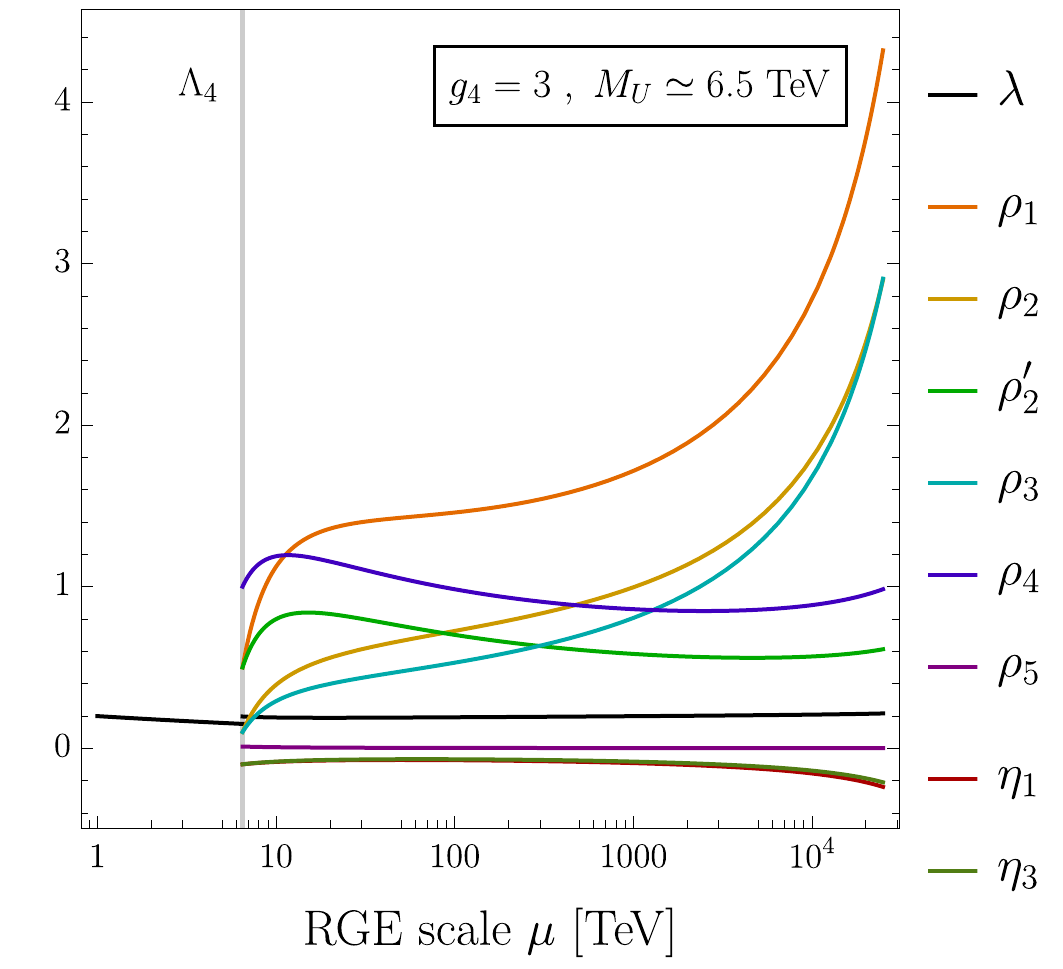}
  \end{subfigure}
  \caption{RG running of the gauge and top Yukawa couplings (left) and quartic couplings of the scalar potential (right) for the benchmark presented in Table~\ref{tbl:benchmark}.}
  \label{fig:RGEs}
\end{figure}

In Table~\ref{tbl:benchmark} we give a benchmark that satisfies the IR boundary conditions and in Figure~\ref{fig:RGEs} we show the resulting RG evolution of the couplings. The fit to the $B$-physics anomalies and the collider bounds on the 4321 gauge bosons prefer a large $g_4$ coupling i.e. $g_4 (\Lambda_4) \gtrsim 2$. On the other hand, the experimental bounds on the scalar masses are marginally surpassed by taking the couplings of the scalar sector within the range $\mathcal{O}(0.1-1)$. The only exception is the coupling $\rho_5$ which mainly fixes the mass of $S_0$ and therefore can be kept one order of magnitude smaller. In fact, as we argued in Section~\ref{sec:scalar_sector}, the smallness of this parameter is technically natural. In fact, larger values for $\rho_5$ would imply also larger values for the rest of the $\rho_i$ parameters in order to keep the $u$ parameter of Eq.~\eqref{eq:u_par} real i.e. $u^2>0$, and as we will see next, this severely restricts the extrapolation the 4321 model in the UV.

 \begin{table}[!ht]
\centering
\begin{center}\renewcommand\arraystretch{1.3}
\begin{tabular}{||c|c||c|c||}
\hline \hline
$\Omega$ VEVs               & values & scalar couplings                              & values \\ \hline
$\omega_1 $      &  $1.5~\rm TeV$      & $\lambda$  &  0.2      \\
$\omega_3 $      &   $4~\rm TeV$     & $\rho_1$   &   0.5     \\ \cline{1-2}
mass parameters               &      &    $\rho_2$   &   0.1         \\ \cline{1-2}
$\mu_H^2$ &   $(1.6~\rm{TeV})^2$     &  $\rho_2'$    &   0.5     \\
$m_{\Omega_1}^2$  &   $-(1.8~\rm{TeV})^2$     &  $\rho_3$    &    0.1    \\
$m_{\Omega_3}^2$      &  $-(2.5~\rm{TeV})^2$   &   $\rho_4$      &   1   \\ \cline{1-2}
gauge coupling              &        &  $\rho_5$   &   0.01     \\ \cline{1-2}
\multirow{2}{*}{$g_4$}            &   \multirow{2}{*}{$3$}      &  $\eta_1$   &  -0.1      \\ 
                              &        &  $\eta_3$    &  -0.1      \\    \hline \hline
\end{tabular}
 \end{center}
\caption{Benchmark for the boundary conditions of the model parameters at scale $\Lambda_4$.}
\label{tbl:benchmark}
\end{table}

Even though $g_4$ is an asymptotically free gauge coupling, the running is slow and its value remains $\mathcal {O}(1)$ even at high energies. Consequently, since the $\beta$-functions of the various parameters are explicitly or indirectly dependent on $g_4$, the whole RG evolution is effectively driven by it. In particular, $\beta(\rho_1)$ contains a term, which solely depends on $g_4^4$ and has the largest numerical prefactor in comparison to other similar terms appearing in the rest of the $\beta$-functions (see Eq.~\eqref{eq:betarho1}).
The reason is that the purely gauge correction to the $\Omega_1$ quartic term (box diagram) includes all possible contractions of $SU(4)$, contrary for example to the $\Omega_3$ quartic for which the different contractions are split between the two terms with couplings $\rho_2$ and $\rho_2'$. As a result $\rho_1$ is a monotonically increasing function of the energy scale $\mu$ and exhibits the fastest running among the couplings leading to a Landau pole.

\begin{figure}
  \centering
	\includegraphics[width=0.6\textwidth]{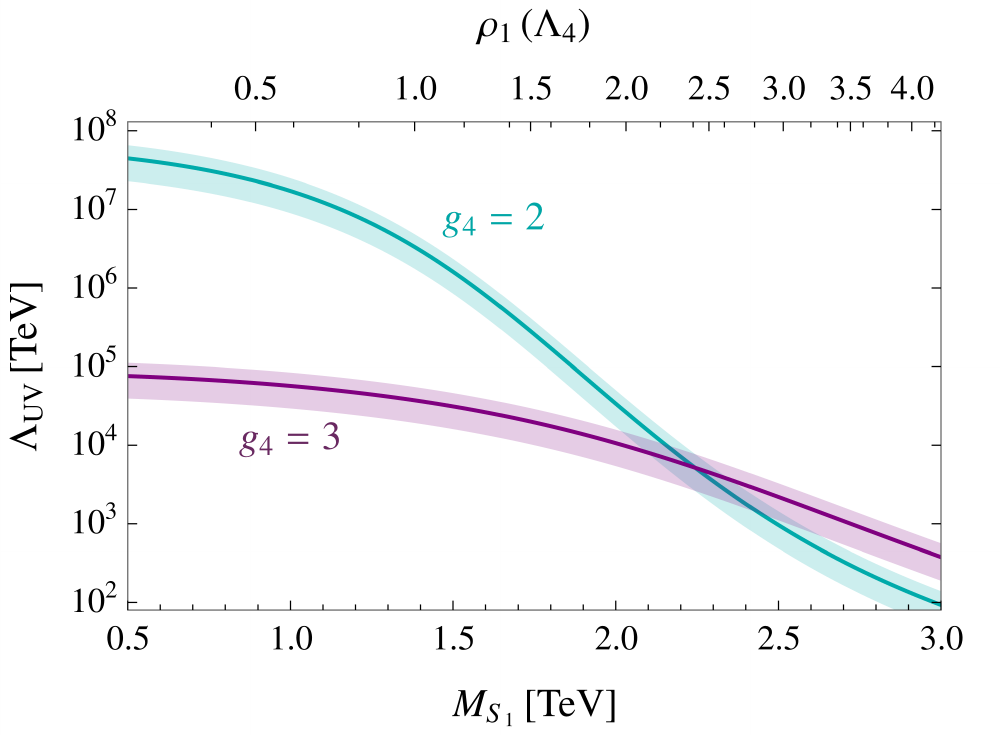}
  \caption{Dependence of the cut-off scale of the 4321 model as a function of the mass of the $S_1$ radial. The mass of $S_1$ is mostly governed by the value of $\rho_1$ at $\Lambda_4$, shown in the upper horizontal axis. The bands corresponds to a $50\%$ variation around the matching scale, set at central value $\Lambda_4=M_U$.} 
  \label{fig:landau_poles}
\end{figure}

Depending on the values of $g_4$ and $\rho_1$ at $\Lambda_4$ we can estimate the energy scale where the scalar couplings grow larger than $\sim 4\pi$ and where the theory becomes non-perturbative. In Figure~\ref{fig:landau_poles} we plot this scale as a function of the mass of the radial $S_1$, which is the scalar mass that is most sensitive to $\rho_1(\Lambda_4)$.
For the benchmark values in Table \ref{tbl:benchmark} and the minimum requirement of radials heavier than $500~\rm GeV$, we find that the perturbativity range of the theory for $g_4=2$ ($M_U\approx 4.5$ TeV) reaches almost $\mathcal{O}(10^8~\rm TeV)$, while for $g_4=3$ ($M_U\approx 6.5$ TeV) it only reaches $\mathcal{O}(10^5~\rm TeV)$. Furthermore, we see that if we require heavier radials this range drastically drops. Notice that the $g_2=3$ curve falls slower than the $g_4=2$ curve for higher $\rho_1$ values, because the  $-\frac{45}{2}g_4^2\rho_1$ term in Eq. \eqref{eq:betarho1} starts to dominate.
Characteristically, the Landau poles appear at even lower energies when one further requires all the radial scalar states heavier than $2~\rm{TeV}$. In this case, couplings blow up at 
$\mathcal{O}(100~\text{TeV})$.  
Considering the above, we argue that the limit of degenerate radials or radials with masses equal or heavier than the leptoquark, which has been used in the literature for simplicity (see for example Refs.~\cite{Fuentes-Martin:2019ign,Fuentes-Martin:2020luw,Fuentes-Martin:2020hvc,Cornella:2021sby}), is not favored. Notice also that our results are relevant to all current versions of 4321 models and consequently impose generic upper bounds for the cut-off scale of 4321 model. In Section~\ref{sec:UV_conditions} we will see that the indicated scale for new physics is even lower if one wishes to unify in the UV.

In this context of fundamental scalars, the question can be posed whether new degrees of freedom can appear at intermediate energies to tame this effect by generating appropriate (possibly fine-tuned) cancellations in the $\beta$-functions. This, however, seems difficult to arrange in our new physics setup (see Appendix \ref{app:subleading_contributions}). The reason is that in order to cancel the $\frac{99}{16}g_4^4$ term in Eq.~\eqref{eq:betarho1}, besides having the opposite sign, a cancellation term must either have a coupling as large as $g_4$ or for a coupling less than 1 an unrealistically large prefactor of $\sim 500(100)$ for $g_4=3(2)$. In the former case, couplings such as Yukawa or scalar couplings, if large,  would quickly develop Landau poles themselves. Thus, it is desirable that the large couplings in the IR be asymptotically free. However in the simplest case where they are e.g. gauge couplings, the breaking of the additional gauge symmetry would require the presence of even more scalars and scalar couplings with similar issues in their RG running. As a result, a solution within this context, if at all feasible, would probably be highly fine-tuned.

\subsection{Radiative electroweak symmetry breaking}
\label{sec:REWSB}

\begin{figure}[H]
    \centering
	\includegraphics[width=.82\textwidth]{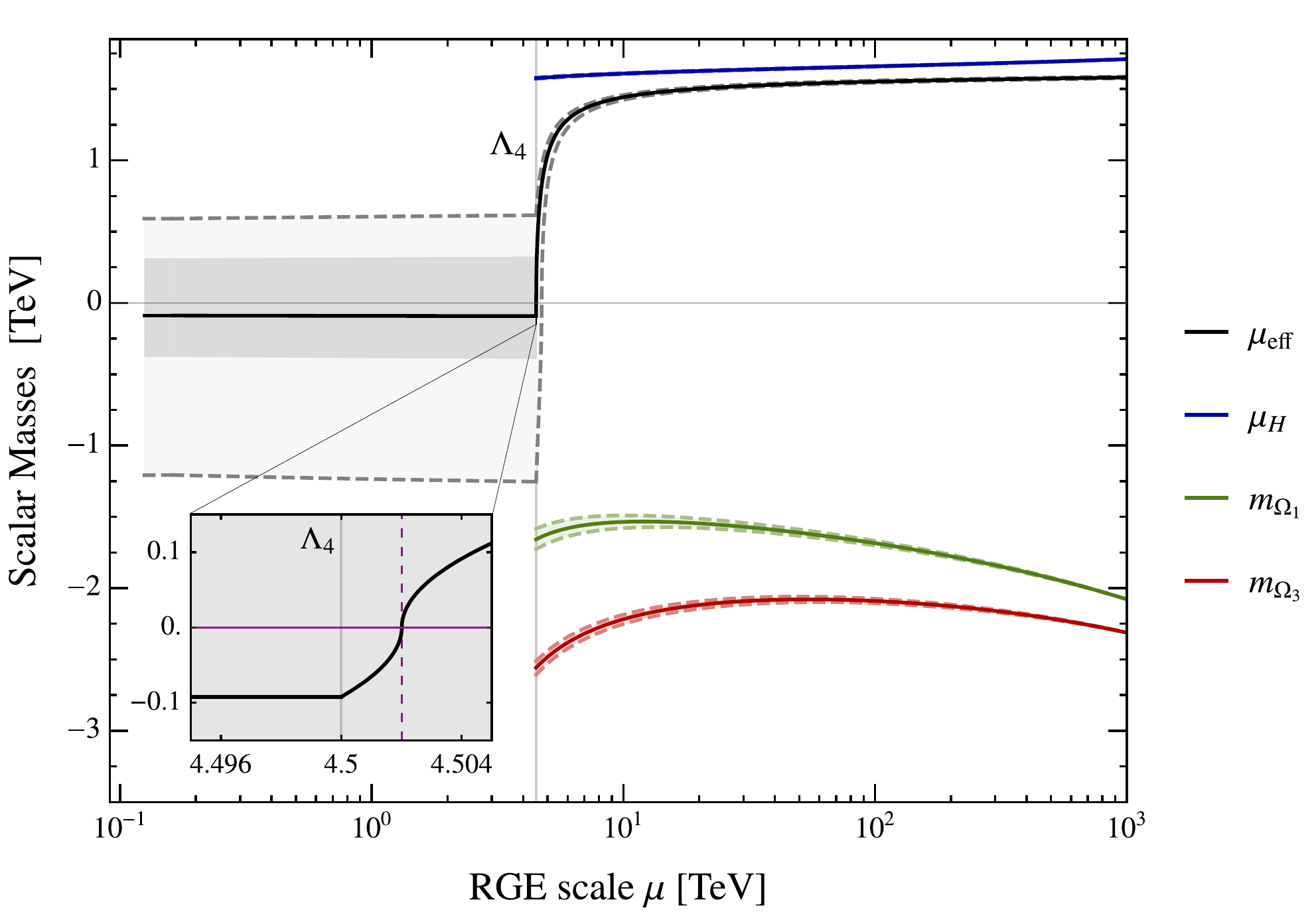}
  \caption{The RG evolution of the scalar sector mass squared parameters $\mu_H^2$, $m_{\Omega_1}^2,$ $m_{\Omega_3}^2$, and $\mu_\text{eff}^2$. For each, the quantity plotted is defined as $X \equiv \text{sign}(X^2) |X|$ ($X = \mu_H, m_{\Omega_1}, m_{\Omega_3}, \mu_\text{eff}$).
  The inset plot focuses on where $\mu_\text{eff}$ passes through $0$, radiatively triggering EWSB near $\Lambda_4 = 4.5$ TeV. We also show the effect of individually varying each quartic coupling as $|\delta (\lambda, \eta, \rho)_i| \leq  5\%$ at $\Lambda_\text{UV}$. This variation propagates along the RG flow of the mass squared parameters, represented by a band enclosed by dashed lines. We also overlay the effect on $\mu_\text{eff}$ from the smaller variations $\delta (\lambda, \eta, \rho)_i$ in the $\pm1\%$ range with a dark gray band.\protect\footnotemark
  }
  \label{fig:massRGE-all}
\end{figure}

Another feature emerges when studying the RG behavior of the 4321 model above $\Lambda_4$: EWSB triggered by radiative effects.
Spontaneous EWSB requires $\mu_\text{eff}^2 < 0$, where $\mu^2_\text{eff}$ is defined in Eq.~(\ref{eq:effectivemass}). For the 4321 model benchmark given in Table~\ref{tbl:benchmark}, $\mu_\text{eff}^2$ flows from positive in the UV to negative in the IR, radiatively triggering EWSB as shown in Figure~\ref{fig:massRGE-all}.

\footnotetext{We thank Ulrich Haisch for pointing out that the perturbative determination of $\mu_{\rm eff}$ in Eq.~\eqref{eq:effectivemass} breaks down at  energies much higher than the matching scale. Therefore, the black curve in Fig.~\ref{fig:massRGE-all} cannot be trusted in the utmost right part of the figure. Nevertheless, we expect it to be reliable in the region where radiative EWSB happens, \textit{i.e.} in the window of the inset plot.}

To see how radiative EWSB occurs, consider the limit $\eta_{1, 3}\to0$, so that the Higgs is decoupled from $\Omega_{1, 3}$. In that limit, $\mu_\text{eff}^2 = \mu_H^2$, and the RG flow of $\mu_H^2$ is controlled by its $\beta$-function given in Eq.~(\ref{eq:betamH1}). In the absence of the $\Omega_{1, 3}$ parameters, the one-loop $\beta$-function is proportional to $\mu_H^2$ and thus a $\mu_H^2$ sign flip from radiative effects alone is not possible. Instead, as $\mu_H$ approaches $0$, it also approaches an approximate fixed point, and $\beta_{\mu_H^2}$ vanishes. Consequently, we see that the coupling of the Higgs to the additional scalars is primarily responsible for triggering EWSB radiatively, as has been pointed out in~\cite{Babu:2016gpg}. In~\cite{Babu:2016gpg}, it was argued that the new scalar states must couple to the Higgs with positive mixing couplings in order to ensure $\mu_H$ flows to negative values in the IR. Here since our $\Omega_{1,3}$ masses are negative (for 4321 spontaneous symmetry breaking), we choose $\eta_{1, 3}<0$ to reproduce the same behavior. 

Note that once $\eta_{1,3}\neq 0 $, however, the relevant parameter to track over the RG flow is $\mu_\text{eff}^2$ rather than $\mu_H$, and its running is a complicated combination of radiative contributions to $\mu_H^2, m_{\Omega_1}^2,$ and $m_{\Omega_3}^2$, as one can infer from Eq.~(\ref{eq:effectivemass}). In this case, it is not clear that $\eta_{1,3}$ are the dominant contributions to radiative EWSB. Instead, radiative EWSB will be controlled not just by the mixing couplings $\eta_{1, 3}$, but also by the $\rho_i$ self-couplings in the $\Omega_{1, 3}$ system, see
Eqs.~(\ref{eq:Omegapotential},~\ref{eq:betamOmega1}-\ref{eq:betamOmega3}, ~\ref{eq:betarho1}-\ref{eq:betaeta1}, ~\ref{eq:betaeta3}). For the benchmark used in Fig.~\ref{fig:massRGE-all}, the $\eta_3, \rho_{1,2,3,4}, \rho_2'$ all contribute significantly to EWSB breaking in the IR. The coupling $\eta_1$ has a smaller impact due to the fact that $\Omega_1$ has fewer degrees of freedom than $\Omega_3$. 

Figure~\ref{fig:massRGE-all} shows that for this benchmark, there is a close coincidence of scales between $\Lambda_4$ and the scale at which $\mu_\text{eff}^2$ changes signs. This is due to the scalar mass mixing that defines $\mu_\text{eff}^2$, equivalent to accounting for the tree-level threshold effects near the $\mathcal{G}_{4321}\to \mathcal{G}_\text{SM}$ breaking scale~\cite{Elias-Miro:2012eoi}. As explained in Section~\ref{sec:ewsb}, in the presence of $\Omega_{1, 3}$ scalar fields, EWSB is a collective effect of the full set of scalar fields, and so as the RG evolution scale $\mu$ approaches $\Lambda_4$, the onset of $\vev{\Omega_{1,3}}\neq0$ is contributing to the emergence of EWSB.

\begin{figure}
  \centering
	\includegraphics[width=.49\textwidth]{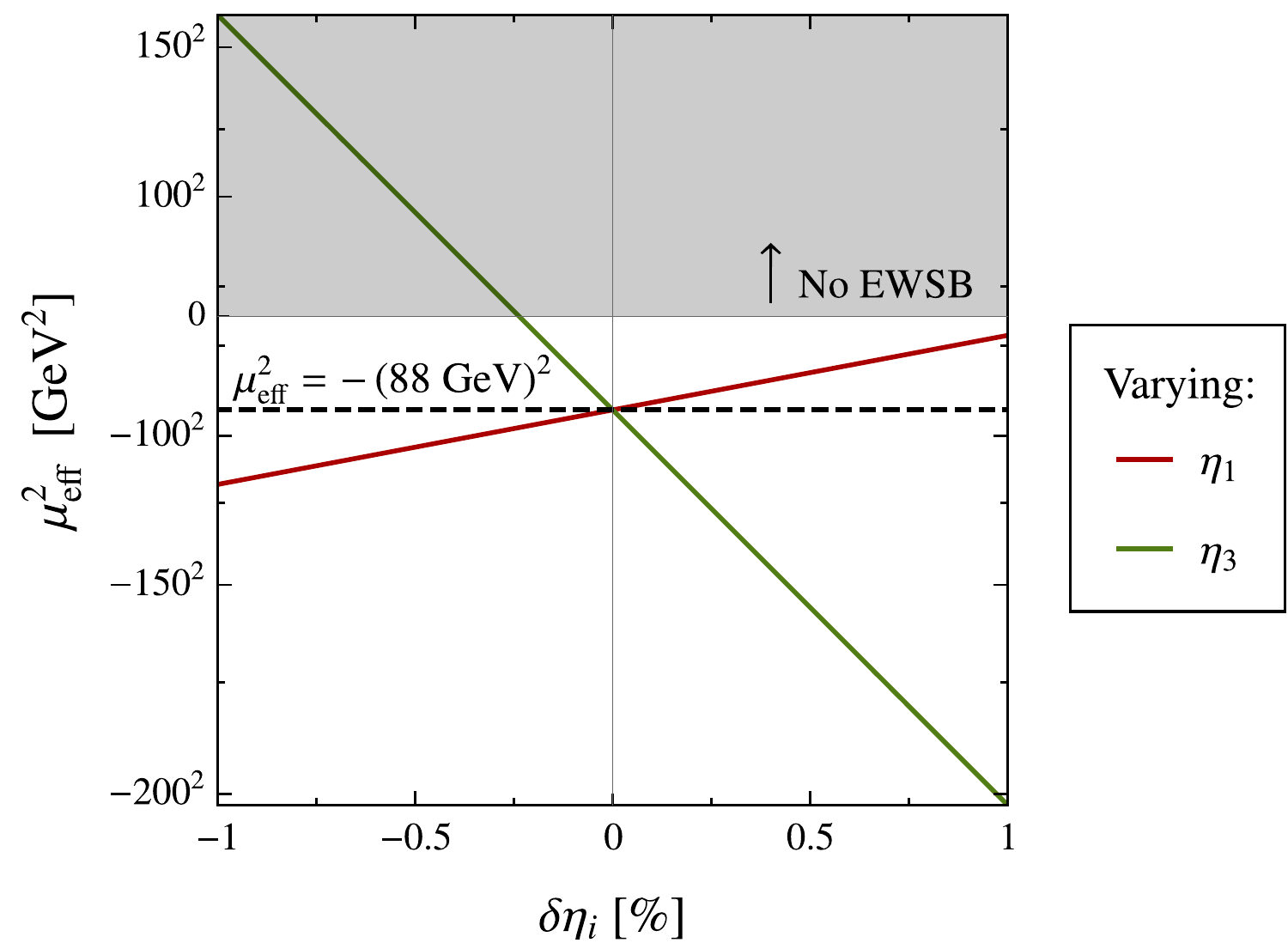}
	\hfill
	\includegraphics[width=.49\textwidth]{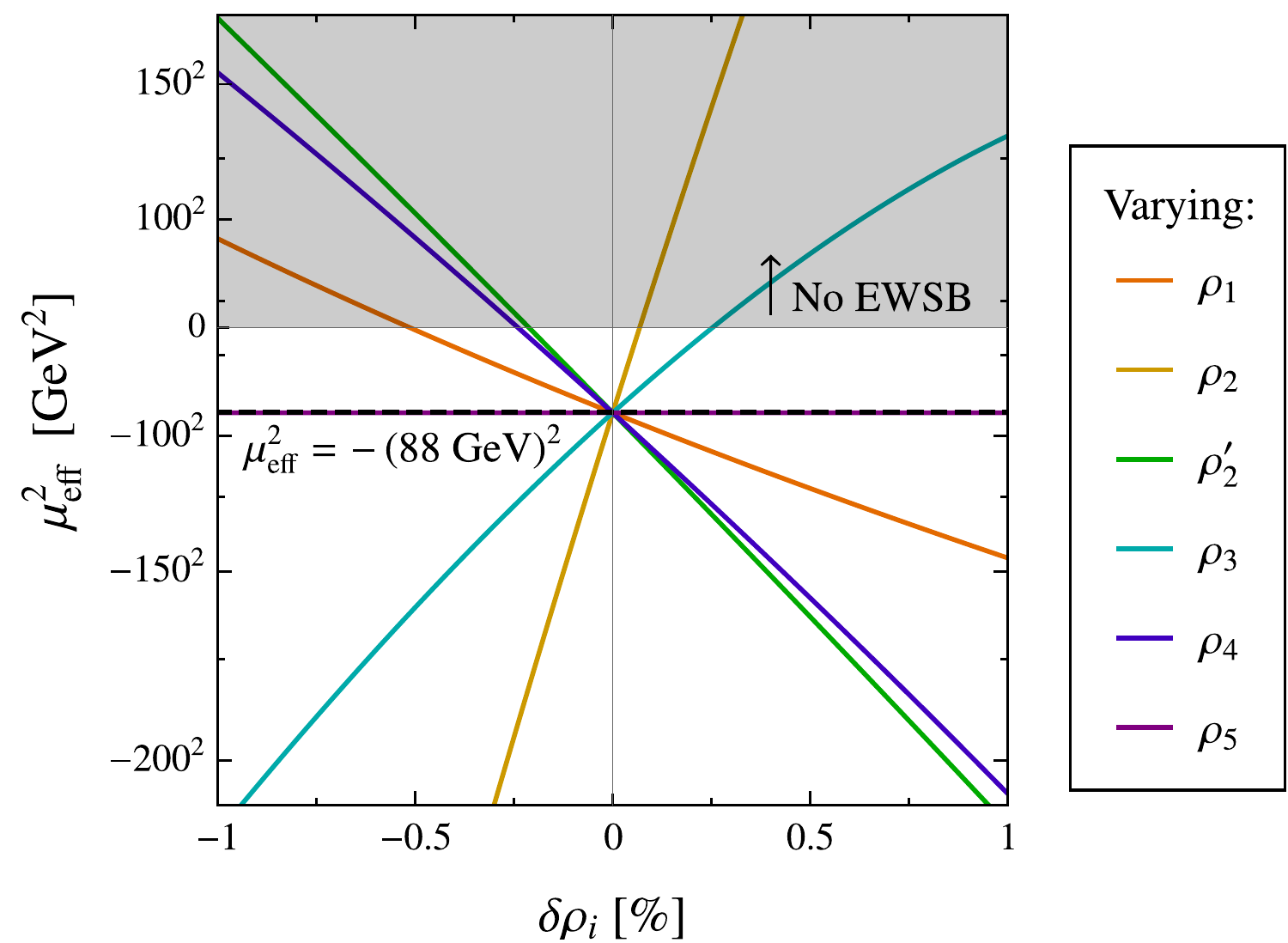}
  \caption{Here we show the sensitivity of  $\mu_\text{eff}^2$ (evaluated at the IR scale $\mu=m_{h_{\rm phys}}$) to small deviations in the Higgs-$\Omega_{1, 3}$ mixing couplings (left) and the $\Omega_{1, 3}$ quartic couplings (right) at $\Lambda_\text{UV}=10^3$ TeV.
  For reference, the value of $\mu_\text{eff}^2$ that reproduces the measured Higgs mass is shown by a dashed, black line. The region where $\mu_\text{eff}^2$ fails to radiatively trigger EWSB is shaded in gray. }
  \label{fig:couplingsFT}
\end{figure}

The steep decrease of $\mu_\text{eff}^2$ from positive to negative values near  $\Lambda_4$ signals a careful arrangement of parameters required in the UV to reproduce the measured Higgs mass in the IR. The Higgs mass parameter $\mu_\text{eff}$ in the IR is indeed sensitive to small deviations of parameters in the UV, as demonstrated in Figure~\ref{fig:couplingsFT}. We present the response of $\mu_\text{eff}^2$ in the IR to varying the quartic couplings at $\Lambda_\text{UV}$ according to
\begin{align}  
    \delta \eta_i 
        &= \frac{ \eta_i - \eta_i^0 }{ \eta_i^0} \ ,
    \quad
    \delta \rho_i 
        = \frac{ \rho_i - \rho_i^0 }{ \rho_i^0}\ , 
\end{align}
where $\eta_i^0, \rho_i^0$ are the benchmark UV couplings values given in Table~\ref{tbl:benchmark}. 
One can employ one of the traditional measures~\cite{Barbieri:1987fn} of fine-tuning $\frac{100}{\Delta_{\rm FT}}\%$, defined as
\begin{equation}
    \Delta_{\rm FT}=\underset{y_i = \{\lambda, \eta, \rho\}_i}{\max} \left|\frac{\delta \mu_\text{eff}^2}{\delta y_i}\right|~, \quad \delta \mu_\text{eff}^2 = \frac{\mu_\text{eff}^2(y_i)-\mu_\text{eff}^2(y^0_i)}{\mu_\text{eff}^2(y^0_i)}~,
\end{equation}
where $\mu_\text{eff}^2(y_i)$ is evaluated at the IR boundary $\mu=m_{h_{\rm phys}}$. The maximum fine-tuning is at the sub-percent level $\Delta_{\rm FT}=\mathcal{O}(10^3)$. In every coupling except $\rho_5$,\footnote{The coupling $\rho_5$ has been chosen to be much smaller than the other couplings for this benchmark, removing $\mu_\text{eff}$'s sensitivity to $\rho_5$.} variations of $\mathcal{O}(1\%)$ can yield positive $\mu_\text{eff}^2$ values in the IR, spoiling EWSB altogether, as shown in the shaded regions of Figure~\ref{fig:couplingsFT}.

The variations at $\Lambda_\text{UV}$ propagate over the RG evolution of the theory, leading to deviations in $\mu_H^2$ and $m_{\Omega_{1, 3}}^2$ at $\Lambda_4$, denoted by the dashed bands in Figure~\ref{fig:massRGE-all}. After the spontaneous symmetry breaking at $\Lambda_4$, deviations in the mass parameters lead to a spread in possible $\mu_\text{eff}^2$ values in the IR, represented by a gray band in Figure~\ref{fig:massRGE-all}. 

Returning to the limit of $\eta_{1,3}\to0$ where the scalars $\Omega_{1, 3}$ are decoupled, we find 
this tuning problem is postponed to the UV. As mentioned above, in this limit we cannot rely on radiative effects to trigger EWSB, and instead  $\mu_\text{eff}^2 \approx - \left( 88 \text{ GeV}\right)^2$ must be arranged directly at $\Lambda_\text{UV}$. A mass hierarchy between $\mu_\text{eff}^2$ and $m_{\Omega_{1, 3}}^2 \sim \text{ TeV}^2$ will endure throughout the flow from the UV to IR. This imposed hierachy of scales at $\Lambda_\text{UV}$ is the trade-off for alleviating $\mu_\text{eff}$'s sensitivity to the initial conditions $\eta_i (\Lambda_\text{UV}),\ \rho_i(\Lambda_\text{UV})$. The tuning apparent in the sensitive dependence of both the Higgs mass and the radiative EWSB mechanism on couplings set at $\Lambda_\text{UV}$ is thus rooted in the little hierarchy problem, as discussed in~\cite{Allwicher:2020esa}.

\subsection{Unification in the ultraviolet}
\label{sec:UV_conditions}

In this section, we examine the implications of UV constraints on the RG evolution. The constraints we employ are inspired by the UV constructions presented in Refs.~\cite{Bordone:2017bld,King:2021jeo}, in which the 4321 model is seen as an effective theory whose gauge group is embedded in multiple copies of the PS group. The scale $\Lambda_{\rm UV}$ is then the scale where the symmetry group of the UV theory breaks spontaneously to the 4321 gauge group. We also choose this scale as the matching scale between the two theories (for more details see Appendix~\ref{app:UV_embedding}). In those UV embeddings the scalar sectors are considerably more complex. However, they feature two common properties: 
\begin{enumerate}[i)]
    \item The fields $\Omega_{1,3}$ unify in the same multiplet. As derived in Appendix~\ref{app:scalar_UV}, this implies that the following relations should hold at the matching scale $\Lambda_{\rm UV}$:
        \begin{equation}\label{eq:omega_UV}
        m_{\Omega_1}=m_{\Omega_3}~, \qquad
        \rho_1=\rho_2 + \rho_2'~, \quad \rho_2=\rho_3~, \quad  \rho_2'=\rho_4~. 
        \end{equation}
    \item A second Higgs doublet with 4321 quantum numbers $H_2 \sim (\mathbf{1},\mathbf{1},\mathbf{2})_{-1/2}$ and mass parameter $\mu_2$ is present. The Higgs potential is that of a Two Higgs Doublet Model (2HDM) as written in Eq.~\eqref{eq:2HDMpotential} and the mixing  potential is given by Eq.~\eqref{eq:portal-2HDMpotential}. The UV constraints derived in Appendix~\ref{app:scalar_UV} give:
    \begin{equation} \label{eq:higgs_UV}
     \mu_H=\mu_2~, \qquad \lambda= \lambda_2 = \lambda_3~, \qquad \eta_1=\eta_2= \eta_3 = \eta_4~.
    \end{equation}
\end{enumerate}   

The 2HDM slightly alters the procedure outlined in the beginning of the section. First of all, we assume a simplified 2HDM potential \eqref{eq:2HDMpotential} in the Higgs basis i.e. $\mu_2^2$ remains positive at all energies and the $H_2$ field does not take a VEV. However, in this case, Eq.~\eqref{eq:higgs_UV} implies that radiative EWSB must happen in order to generate a negative $\mu_{\rm eff}^2$.  Then collider bounds can be simply avoided by requiring that the masses of the heavy Higgs radials are above $500~\rm GeV$.\footnote{Notice that this bound is sufficient also for other types of 2HDMs, e.g. where flavor constraints are relevant~\cite{Haller:2018nnx}.} Regardless of the choice of $\lambda_i$, this constraint is readily satisfied for $\mu_2^2 (500~\rm GeV) \gtrsim (500~\rm GeV)^2$ (see Eq.~\eqref{eq:2HDM_spectrum}). Moreover, the $\lambda_i$ must satisfy the BFB conditions for the 2HDM (see Eq.~\eqref{eq:2HDM_BFB}).

Here, we solve the RG equations by using the full $\beta$-functions listed in Appendix~\ref{app:4321_beta}. From the scale $\Lambda_{4}$ down to the scale $\mu_2$, we use the $\beta$-functions of the 2HDM (see Appendix~\ref{app:2HDM_beta}) and at $\mu_2$ the degenerate heavy Higgs radials are collectively integrated out. Because $\mu_2$ is much closer to the electroweak scale, the threshold corrections in this case are much smaller and can be omitted. On the other hand, threshold corrections at $\Lambda_{\rm UV}$ could originate from different scales and induce important deviations to the expectations Eq.~\eqref{eq:omega_UV} and Eq.~\eqref{eq:higgs_UV}. We quantify this uncertainty by allowing deviations up to 10 \% from the exact equalities. Despite that, due to the fast running of certain parameters, it is not  a priori clear whether the RG evolution can be reconciled with even approximate conditions.

Our goal becomes then to search for sets of initial parameter values that satisfies the IR boundary conditions and via their RG evolution yield sets of values that approximately satisfy the UV boundary conditions. To obtain such sets, we perform a numerical scan over the parameter space by solving the RG equations for $\mathcal O(10^5)$ random sets for benchmarks $g_4=2,3$ and $\Lambda_{\rm UV}=10,10^{2},10^{3},10^{4},10^{5}~\rm TeV$. In order to increase the efficiency of the scan in this high-dimensional space, we employ a Markov Chain Monte Carlo algorithm (Hastings-Metropolis) for the generation of trial sets. Additionally, we require that the UV values of the couplings are still perturbative. It is worth mentioning that the non-trivial part of the search concerns only the $\rho$ and $m_{\Omega_{1,3}}$ parameters. This is because in the absence of concrete IR conditions for $\eta$, $\lambda_{2,3}$, $\mu_2$ and a loose dependence on $g_4$, one can always adjust the IR values in order to find appropriate UV ones.\footnote{We note that in multi-site UV constructions, like the $\rm PS^3$ model, the $\Omega$ and the Higgs fields may unify into their respective multiplets at different scales. Because, as mentioned, the problem of matching to the conditions Eq.~\eqref{eq:higgs_UV} is much less constrained, one could also achieve this without significant deviations from our discussion.}

\begin{figure}[H]
  \begin{subfigure}{0.43\textwidth}
  \includegraphics[width=\textwidth]{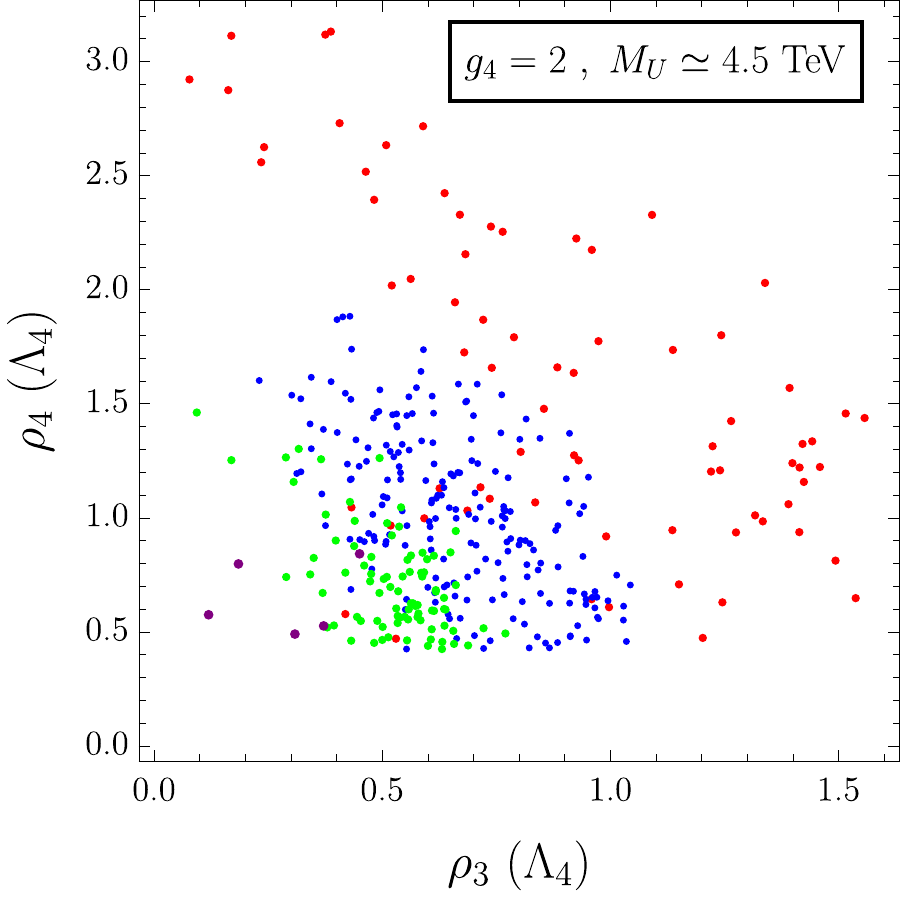}   
    \end{subfigure} 
    \hspace{0.1cm}
    \begin{subfigure}{0.43\textwidth}
	\includegraphics[width=\textwidth]{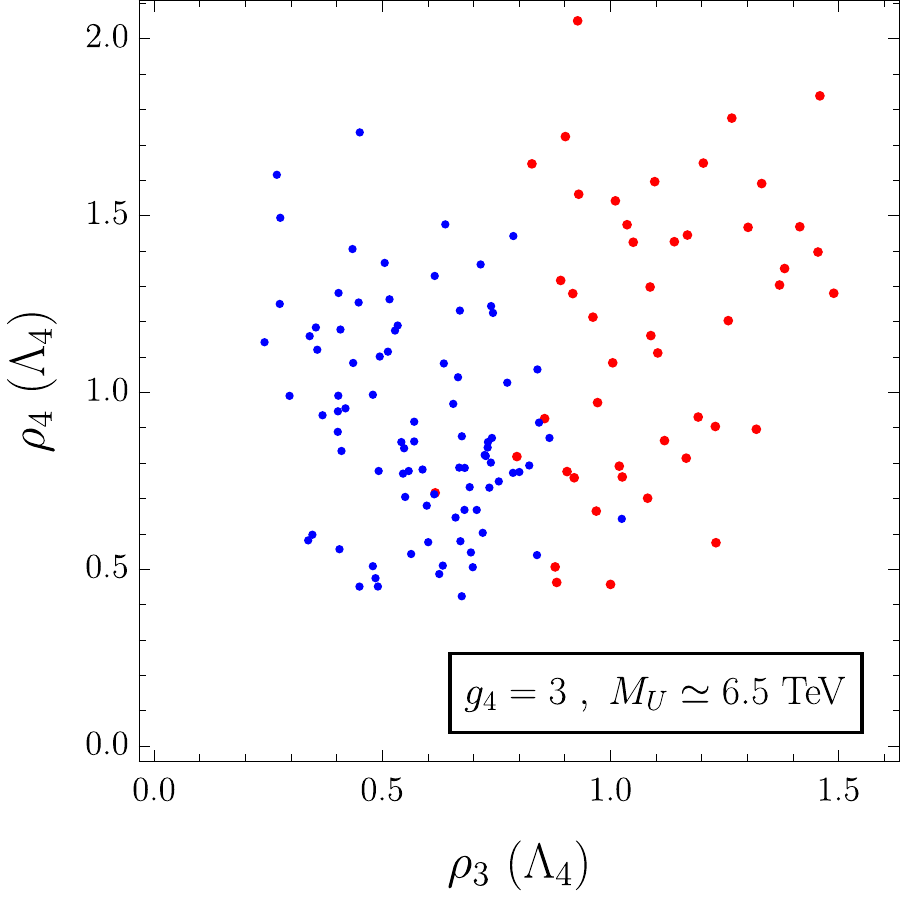}
  \end{subfigure}
  \hspace{0.1cm}
   \begin{subfigure}{0.08\textwidth}
   \vspace{-1cm}
   \centering
	\includegraphics[width=\textwidth]{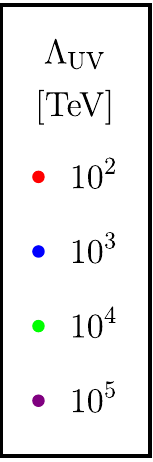}
  \end{subfigure}
  \caption{Benchmark points in the $\rho_3(\Lambda_4)$ -- $\rho_4(\Lambda_4)$ plane satisfying the UV conditions Eqs.~\eqref{eq:omega_UV} and \eqref{eq:higgs_UV} for different $g_4$ couplings and unification scales $\Lambda_{\text{UV}}$.
  }
  \label{fig:scan}
\end{figure}

In Figure~\ref{fig:scan} we present the distribution of the obtained initial values of a representative pair of parameters, namely $\rho_3$ and $\rho_4$. The first observation is that we find no acceptable sets for unification scales as low as $10 ~\rm TeV$ and very few sets for scales as high as $10^{5} ~\rm TeV$ (and only in the case of $g_4=2$). This is explained by the fact that the different $\rho$ parameters exhibit different running behaviors and thus convergence to the UV conditions is possible only in the intermediate scales $10^{3}-10^{4}~\rm TeV$. For instance, the relation $\rho_1=\rho_2+\rho_2'$ cannot be satisfied at higher energies due to the faster running of $\rho_1$ and the fact that $\rho_2'$ is monotonically decreasing as opposed to the other two. In total, we found more sets for smaller $g_4$ because, as already discussed, the overall running in this case is slower and thus more suitable for achieving unification.

\begin{figure}[H]
  \begin{subfigure}{0.5\textwidth}
  \includegraphics[width=\textwidth]{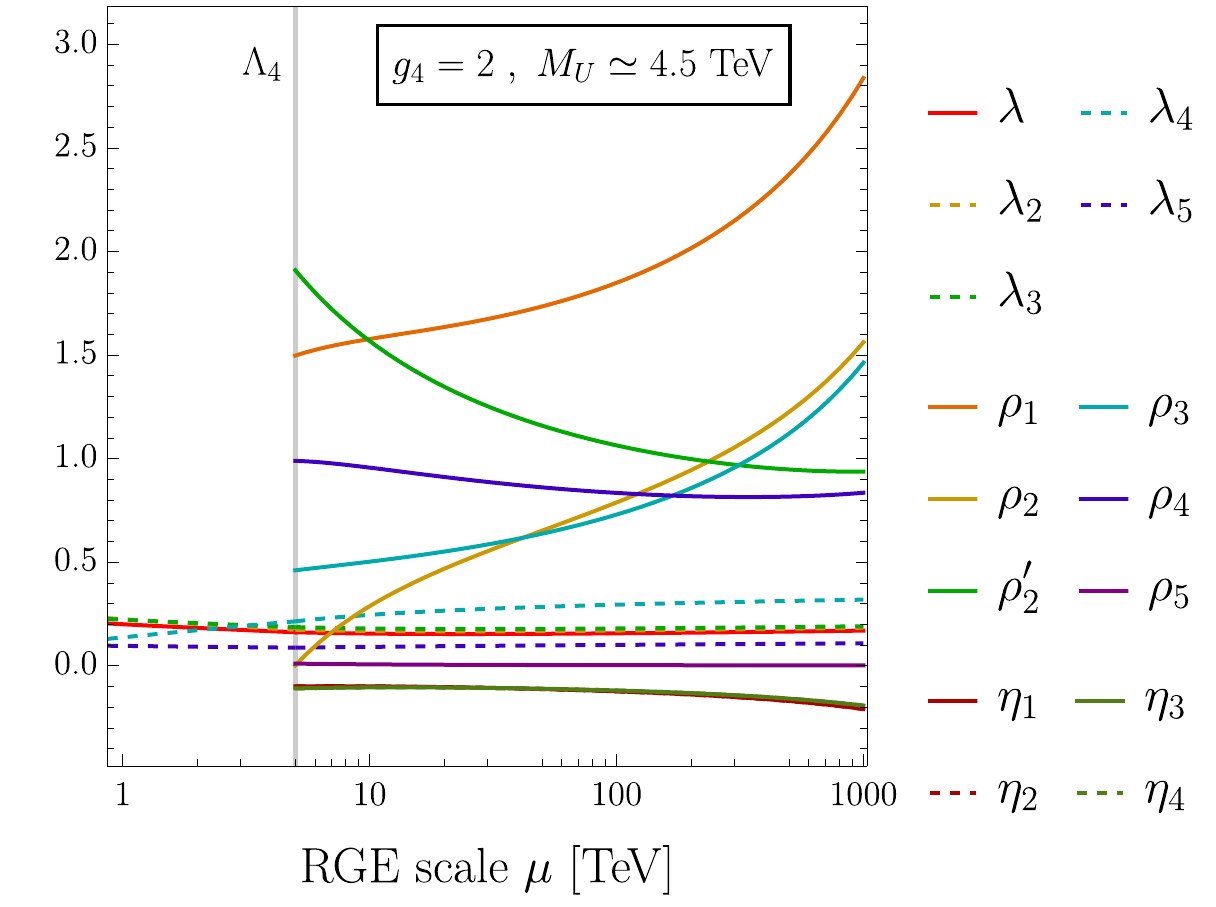}     
    \end{subfigure}
    \begin{subfigure}{0.5\textwidth}
	\includegraphics[width=\textwidth]{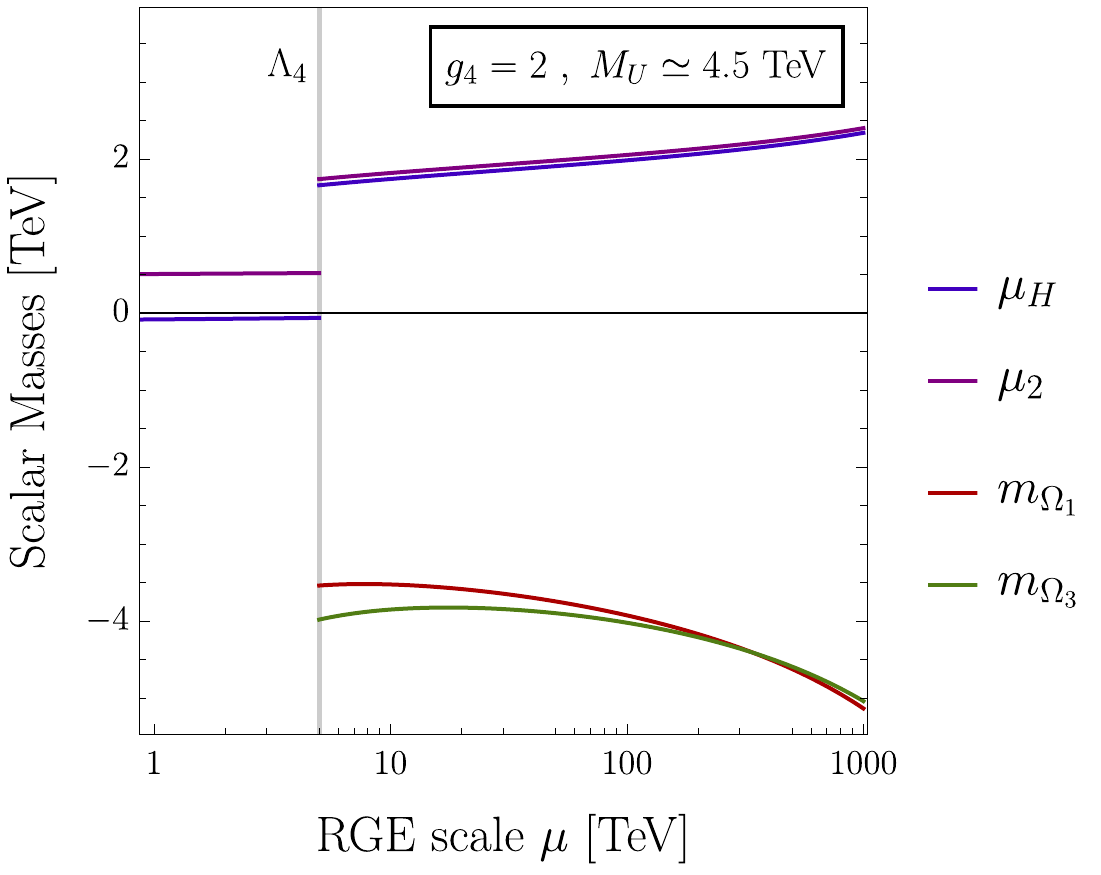}
  \end{subfigure}
  \caption{RG running of the gauge and top Yukawa couplings (left) and couplings of the scalar potential (right) with a benchmark reaching the UV conditions Eqs.~\eqref{eq:omega_UV} and \eqref{eq:higgs_UV}. The y-axis, scalar masses, is defined as in Figure \ref{fig:massRGE-all}.
  }
  \label{fig:RGE_UV}
\end{figure}

On the other hand, for lower unification scales the RG evolution spans a shorter energy range and the running has to be accelerated in order to meet the UV conditions. This is the reason why in Figure~\ref{fig:scan} we see that the initial values of the parameters increase as $\Lambda_{\rm UV}$ decreases so that the terms in the $\beta$-functions proportional to the couplings themselves are enhanced. Altogether, the attempt to unify the couplings requires $\Lambda_{\rm UV}$ approximately three orders of magnitude lower than the cut-off scales discussed in Section~\ref{sec:perturbativity} for the generic case. We conclude the discussion by illustrating in Figure~\ref{fig:RGE_UV} the unification of the masses and couplings using as an example the RG evolution generated by one of the obtained sets. 

\section{Conclusions}
\label{sec:conclusions}
  
In this work, we performed an analysis of the RG evolution of the 4321 model, which revealed several interesting features inherent to the scalar sector of the model. 
We observed that the large value of $g_4$ plays an important role in the overall behavior of the running and it can lead to the appearance of Landau poles even though the gauge coupling itself is asymptotically free. Phenomenological constraints on the $\Omega_{1,3}$ radial modes prevent their quartic couplings from assuming arbitrarily small IR values, leading these quartic couplings to blow up, and compromising the validity of the 4321 model, at energies far below the GUT scale. Additionally, we found that it is possible to satisfy relations from motivated UV embeddings within the energy range $\mathcal{O}(10^2-10^4)~\rm TeV$.

Intriguingly, the emergence of Landau poles seems to be intrinsic to a large class of the models proposed as combined explanations for both charged- and neutral-current $B$-physics anomalies. Depending on the model, the problem may appear in different sectors of the theory but typically involves the presence of fundamental scalars (see Refs.~\cite{DiLuzio:2018zxy,Azatov:2018kzb,Marzocca:2021azj} for examples of Landau poles in the Yukawa couplings). As discussed, it is highly non-trivial to address this problem at the level of $\beta$-functions by introducing additional fields.
This might indicate that the 4321 gauge symmetry is broken not by fundamental scalars, but rather \textit{\`a la technicolor} by the condensate of a strongly coupled sector as in Ref.~\cite{Fuentes-Martin:2020bnh,Fuentes-Martin:2022xnb}.
Another interesting possibility is that the leptoquark arises directly as a composite resonance of a new strongly coupled sector as in Refs.~\cite{Barbieri:2016las,Marzocca:2018wcf}.
For example, the model in Ref.~\cite{Marzocca:2018wcf} with composite scalar leptoquarks and Higgs is free of low-energy Landau poles.

Upon studying the RG flow of the mass squared parameters in the scalar sector, we found that the 4321 model can accommodate radiative EWSB due to the additional scalar fields coupled to the Higgs. While radiative EWSB requires positive contributions to the $\beta$-function generated by the $H$-$\Omega_{1, 3}$ mixing, the IR value of the parameter that controls EWSB, $\mu_\text{eff}^2$, relies on the whole system of scalar quartic couplings, mixing couplings, and $\Omega_{1, 3}$ mass parameters. 
We quantified the sensitivity of the Higgs mass to the UV values of the parameters.
The resulting fine-tuning reflected the little hierarchy problem associated with the new TeV scalars.

Ultimately, if the B-physics anomalies are confirmed as clear signals of new physics, leptoquarks arise as the most plausible explanation. Given that all leptoquark models contain additional scalar particles, we stress the importance of carefully examining their sectors, and in particular, the RG evolution of the scalar couplings. As we have seen in the case of 4321, new features can emerge upon analysis of radiative effects, constraining the possibilities for UV constructions.

\acknowledgments

We thank Gino Isidori for discussion and comments on this work, Ben Stefanek for useful discussion and Sandro M\"achler for crosschecks. 
S.T. acknowledges support by MIUR grant PRIN 2017L5W2PT.
R.H. is supported by the STFC under grant ST/P001246/. 
J.P. acknowledges the European Research Council (ERC) under the European Union’s Horizon 2020 research and innovation programme under grant agreement 833280 (FLAY).

\appendix

\section{$\beta$-functions of the 4321 model}
\label{app:4321_beta}
 
The $\beta$-functions for the gauge couplings at one-loop level are given by the master formula: 
\begin{equation}
\mu\frac{d \alpha_i}{d\mu}
\equiv  \frac{1}{16 \pi^2}\,  \beta_{\alpha_i}
=  -\frac{\alpha_i^2}{2\pi} b_i~,
\end{equation}
where
\begin{equation} \label{eq:beta_couplings_formula}
b_i = \frac{11}{3} C_2(G) - \frac{4}{3} \sum_f \kappa_f S_2(R_f) - \sum_s \frac{1}{6} S_2(R_s)~.
\end{equation}
The first term contains the quadratic Casimir of the gauge boson representation $C_2(G)$, the second and the third the sums over the Dynkin indices for the representations of all the fermions $S_2(R_f)$ ($\kappa_f = 1/2$ or $\kappa_f = 1$ for 2- or 4-component fermions) and the scalars $S_2(R_s)$, respectively. Taking into account the full particle content of the 4321 model 
the resulting values for $b_i$ are
\begin{align}
    \begin{aligned}
 \label{eq:beta_gauge_couplings}
U(1)': b_1 &= -\frac{115}{18}~,& \quad
SU(2)_L: b_2 &= \frac{19}{6}-\frac83 n_{\rm VL}~, ~\\
SU(3)_{1+2}: b_3 &= \frac{23}{3}~,& \quad
SU(4)_3: b_4 &= \frac{38}{3} - \frac43 n_{\rm VL}~.
    \end{aligned}
\end{align}
where $n_{\rm VL}$ is the number of vector-like fermions.

The $\beta$-functions for the couplings of the potential in Eq.~\eqref{eq:Scalar-potential} 
are given at one-loop order:
\begingroup
\allowdisplaybreaks
\begin{align}
 \beta_{\mu_H^2}
	&= \left( 6 \lambda + 8 y_t^2 - \frac32 g_1^2 - \frac92 g_2^2 \right) \mu_H^2 
		+ 8 \eta_1 m_{\Omega_1}^2
		+ 24 \eta_3 m_{\Omega_3}^2
		\gr{\,+\,  ( 4 \lambda_3 + 2 \lambda_4 ) \mu_2^2 }~,
\label{eq:betamH1}
\\
\gr{ \beta_{\mu_2^2}}
	&\gr{\,= \left( 6 \lambda_2 +
	8 y_t^2-\frac32 g_1^2 - \frac92 g_2^2 \right) \mu_2^2 
		+  ( 4 \lambda_3 +2 \lambda_4 ) \mu_H^2 
		+ 8 \eta_2 m_{\Omega_1}^2
		+ 24 \eta_4 m_{\Omega_3}^2~,}
\label{eq:betamH2}
\\
\beta_{ m_{\Omega_1}^2 }
	&= \left( 10 \rho_1  -\frac32 g_1^2 - \frac{45}{4} g_4^2 \right) m_{\Omega_1}^2 
		+ ( 24 \rho_3 + 6 \rho_4 ) m_{\Omega_3}^2
		+ 4 \eta_1 \mu_H^2
		\gr{\,+\, 4 \eta_2 \mu_2^2~},
\label{eq:betamOmega1}
\\
\beta_{ m_{\Omega_3}^2 }
	&= \left(  26 \rho_2 
		+  14 \rho_2'  
		- \frac16 g_1^2 
		- 8 g_3^2
		- \frac{45}{4} g_4^2 \right)  m_{\Omega_3}^2
		+ 4 \eta_3 \mu_H^2
		+ ( 8 \rho_3  +2 \rho_4 ) m_{\Omega_1}^2
		\gr{\,+\, 4 \eta_4 \mu_2^2} ~,
\label{eq:betamOmega3}
\\
\beta_{ \lambda }
	&=\left( 12  \lambda + 16 y_t^2 - 3  g_1^2- 9  g_2^2 \right) \lambda
    	+ 8 \eta_1^2
		+ 24 \eta_3^2
		- 16  y_t^4
	\gr{\,+\, 4 \lambda_3^2 
		+ 4 \lambda_3 \lambda_4 
		+ 2\lambda_4^2 
		+ 2\lambda_5^2}
		\nonumber\\
		&\ \ \ \ \ 
		+\frac34 g_1^4
		+ \frac32 g_1^2 g_2^2
		+ \frac94 g_2^4~,
\label{eq:betalambda1}
\\
\gr{\beta_{ \lambda_2 }}
	&\gr{\,= \left( 12 \lambda_2 + 16 y_t^2 -3 g_1^2 - 9 g_2^2 \right)\lambda_2 
	    + 8 \eta_2^2
		+ 24 \eta_4^2
		- 16  y_t^4
	    + 4 \lambda_3^2 + 4 \lambda_3 \lambda_4 + 2\lambda_4^2 + 2\lambda_5^2}
		\nonumber\\
		&\ \ \ \ \ 
		\gr{+\,\frac34 g_1^4
		+ \frac32 g_1^2 g_2^2
		+ \frac94 g_2^4~,}  
\label{eq:betalambda2}
\\
\gr{\beta_{ \lambda_3 }}
	&\gr{\,= \left(  4 \lambda_3+ 6 \lambda + 6 \lambda_2 - 3 g_1^2
		- 9  g_2^2  + 16  y_t^2  \right) \lambda_3
        + 2 \lambda_4( \lambda + \lambda_2 ) 
		+ 2\lambda_4^2 + 2\lambda_5^2
		- 16  y_t^4  
		}
		\nonumber\\
		&\ \ \ \ \ 
		\gr{
		+\, 8 \eta_1 \eta_2
		+ 24 \eta_3 \eta_4
		+\frac34 g_1^4
		- \frac32 g_1^2 g_2^2
		+ \frac94 g_2^4 
		~,}
\label{eq:betalambda3}
\\
\gr{\beta_{ \lambda_4 }}
	&\gr{\,=   \left( 4 \lambda_4+ 2 \lambda + 2 \lambda_2 + 8 \lambda_3  -3  g_1^2 -9  g_2^2  + 16  y_t^2 \right)  \lambda_4
		+ 8 \lambda_5^2
		+ 3  g_1^2  g_2^2
		+ 16  y_t^4~,}
\label{eq:betalambda4}
\\
\gr{\beta_{ \lambda_5 }}
	&\gr{\,= \left(  2\lambda +  2\lambda_2  + 8 \lambda_3   + 12 \lambda_4 
	+16 y_t^2
	- 9 g_2^2 - 3 g_1^2 \right) \lambda_5~,}
\label{eq:betalambda5}
\\
\label{eq:betarho1}
\beta_{ \rho_1 }
	&= \left( 16 \rho_1
    	- 3 g_1^2 
    	- \frac{45}{2} g_4^2 \right) \rho_1
		+ 24 \rho_3^2
		+ 12 \rho_3 \rho_4
		+ 6 \rho_4^2
		+ 4 \eta_1^2 
		\gr{\,+\, 4 \eta_2^2}
	\nonumber\\
	&\ \ \ \ \ 
		+ \frac34 g_1^4 + \frac94 g_1^2 g_4^2 + \frac{99}{16} g_4^4~, 
\\
\label{eq:betarho2}
\beta_{ \rho_2 }
	&= \left( 32  \rho_2
	    + 28 \rho_2'
	    -\frac13 g_1^2 
	    -16  g_3^2 
	    -\frac{45}{2}  g_4^2\right)   \rho_2
		+ 6 {\rho_2'}^2
		+ 8 \rho_3^2 
		+ 4 \rho_3 \rho_4
		+ 4 \rho_5^2
	\nonumber\\
	&\ \ \ \ \ 
	    + 4 \eta_3^2
		\gr{\,+\, 4 \eta_4^2}
		+\frac{1}{108}g_1^4
		+ \frac{11}{6} g_3^4
		+ \frac{27}{16} g_4^4
		-\frac{1}{9} g_1^2 g_3^2
		-\frac{1}{12} g_1^2g_4^4
		+ \frac{13}{2} g_3^2g_4^2~,		
\\
\label{eq:betarho2p}
\beta_{ \rho_2' }
	&= 
		 \left(14 {\rho_2'} + 12  \rho_2 - \frac13 g_1^2 - 16  g_3^2 - \frac{45 }2 g_4^2 \right)  \rho_2'
		+ 2 \rho_4^2
		- 4 \rho_5^2
        \nonumber\\
		&\ \ \ \ \ 
		+ \frac52 g_3^4
		+ \frac92 g_4^4
		+ \frac13 g_1^2 g_3^2
		+ \frac13 g_1^2 g_4^2
		- \frac72 g_3^2 g_4^2~,
\\
\label{eq:betarho3}
\beta_{ \rho_3 }
	&=  \left( 4 \rho_3 +  10 \rho_1 + 26 \rho_2 + 14 \rho_2' -  \frac53 g_1^2 -8g_3^2 - \frac{45}{2}g_4^2\right)  \rho_3 
	    + 2 \rho_4^2
		+ 4 \rho_5^2
        \nonumber\\
		&\ \ \ \ \ 
		+  2( \rho_1 + 3 \rho_2 +  \rho_2' ) \rho_4 
		+ 4 \eta_1 \eta_3
		\gr{ \,+\, 4 \eta_2 \eta_4 }
		+\frac{1}{12} g_1^4 + \frac{1}{4} g_1^2 g_4^2 + \frac{27}{16} g_4^4~,
\\
\label{eq:betarho4}
\beta_{ \rho_4 }
	&=  \left( 8 \rho_4 + 2 \rho_1 + 2 \rho_2  + 6 \rho_2' + 8\rho_3  -\frac53 g_1^2 -8g_3^2 - \frac{45}{2}g_4^2 \right)  \rho_4
		-4\rho_5^2
        \nonumber\\
		&\ \ \ \ \ 
		-g_1^2 g_4^2
		+\frac92 g_4^4~,
\\
\label{eq:betarho5}
\beta_{ \rho_5 }
	&=  \left[ 6 (\rho_2 -\rho_2' + \rho_3 -\rho_4) 
		-  g_1^2 -12 g_3^2 -\frac{45}2 g_4^2 
		\right]  \rho_5 ~,
\\
\beta_{ \eta_1}
	&=  \left( 4 \eta_1 + 6 \lambda  + 10 \rho_1
	        - 3 g_1^2 - \frac92 g_2^2 - \frac{45}{4}g_4^2 
	        + 8 y_t^2
	        \right) \eta_1
        \nonumber\\
		&\ \ \ \ \ 	
	    \gr{+\, 4 \lambda_3 \eta_2 
	    + 2 \lambda_4 \eta_2}
		+ 24 \rho_3 \eta_3 
		+ 6 \rho_4 \eta_3 
		+\frac34 g_1^4
		~,
\label{eq:betaeta1}
\\
\gr{\beta_{ \eta_2}}
	& \gr{\,= \left(  4 \eta_2 + 6 \lambda_2 + 10 \rho_1 - 3 g_1^2 - \frac92 g_2^2 -             \frac{45}{4}g_4^2 
	    +8 y_t^2 \right) \eta_2 }
        \nonumber\\
		&\ \ \ \ \ 
	    \gr{ +\, 4 \lambda_3 \eta_1 
	     + 2 \lambda_4 \eta_1
		 + 24 \rho_3 \eta_4
		 + 6 \rho_4 \eta_4
		+\frac34 g_1^4~,}
\\
\beta_{ \eta_3 }
	&= \left(  6 \lambda + 26 \rho_2 + 14 \rho_2'  - \frac53 g_1^2 - \frac92  g_2^2 - 8 g_3^2 - \frac{45}{4}g_4^2  
	    +8  y_t^2
	    \right) \eta_3
        \nonumber\\
		&\ \ \ \ \ 
	    \gr{+\,  ( 4 \lambda_3 + 2\lambda_4 ) \eta_4}
		+  ( 8 \rho_3  + 2 \rho_4 )\eta_1 
		+ 4 \eta_3^2
		+ \frac{1}{12} g_1^4
		~,
\label{eq:betaeta3}
\\[1cm]
\gr{\beta_{ \eta_4 }}
	&\gr{\,= 6 \lambda_2 \eta_4
		+  ( 4 \lambda_3 +2 \lambda_4 ) \eta_3
		+ 2 ( 13 \rho_2 + 7 \rho_2') \eta_4
		+  (  8 \rho_3  +2 \rho_4 )\eta_2
		+ 4 \eta_4^2}
	    \nonumber\\
		&\ \ \ \ \ 
	\gr{	- \left( \frac53 g_1^2 + \frac92  g_2^2 + 8 g_3^2 + \frac{45}{4}g_4^2 \right) \eta_4
		+ \frac{1}{12} g_1^4
		+8 \eta_4 Y_{12}~,}
\\
\label{eq:beta_yt}
    \beta_{y_t}
	&=  \left( 
		 \frac{11}{2} y_t^2
		-\frac94 g_2^2
		- \frac{45}{4} g_4^2
		- \frac34 g_1^2 \right)
		y_t~.
\end{align}
\endgroup
where only the top Yukawa coupling was considered from Eq.~\eqref{eq:LYuk}. The terms written in gray describe the running/contribution from a second Higgs doublet in the model (see Appendix~\ref{app:2HDM}). All the $\beta$-functions were crosschecked using the Mathematica package \texttt{RGBeta}~\cite{Thomsen:2021ncy}.

\section{Subleading contributions to the $\beta$-functions}
\label{app:subleading_contributions}

In Section~\ref{sec:scalar_content}, we considered the minimal scalar content for the 4321 model and neglected all Yukawa couplings except for the top. Below, we discuss the influence of additional contributions to the $\beta$-functions, arising from the Yukawa couplings of the vector-like fermions and from the addition of a second Higgs doublet, or other extra scalars.

\subsection{Yukawa sector}
\label{app:Yukawa}

Neglecting the Yukawa couplings of the light generations, the Yukawa sector of the 4321 model reads
 	\begin{align} \label{eq:LYuk}
 	\begin{aligned}
 	-\mathcal{L}_{\rm Yuk} &=
 		y_t\, \bar \Psi_L \tilde H \Psi_R^+ 
 		+ y_b\, \bar \Psi_L  H \Psi_R^- 
 		+ y_+\, \bar \chi_L \tilde H \Psi_R^+ 
 		+ y_-\, \bar \chi_L  H \Psi_R^- \\
 		&+ \lambda_q\,  \bar q_L \Omega_3 \chi_R 
 		+ \lambda_\ell\, \bar \ell_L \Omega_1 \chi_R 
 		+ M_\chi\, \bar \chi_L \chi_R \,  
 		+ \text{h.c.}\,~,
 	\end{aligned}
 	\end{align}	
with $\tilde H= i \sigma_2 H^*$ and where we used the freedom of rotating the multiplet $\begin{pmatrix}
     \Psi_L, \chi_L
\end{pmatrix}$ to remove the mixing term $
\bar \Psi_L\, \chi_R$.
The biggest couplings are $y_t \sim 1$ and $y_+\sim 1/4$, while the other couplings are expected to be smaller, i.e. $\lambda_{q,\ell }\sim 10^{-1}$, $y_b \sim 10^{-2}$ and $y_-\sim 10^{-3}$ from phenomenological considerations \cite{Cornella:2021sby}.
The influence of these terms on the scalar quartics $\beta$-functions were computed and did not yield any relevant contribution. For example, the contribution
\begin{equation} 
    \beta_{\rho_1} \supset -8 |\lambda_\ell|^4 + 8 \rho_1 |\lambda_\ell|^2\,~,
\end{equation}
should be compared to the $g_4^4$ term in Eq.~\eqref{eq:betarho1}.

\subsection{Two higgs doublet model} \label{app:2HDM}
 A second Higgs doublet $H_2  \sim (\boldsymbol{1}, \boldsymbol{1},\boldsymbol{2})_{1/2}$ is theoretically motivated by UV embeddings of 4321. For instance, in the $\rm PS^3$ model~\cite{Bordone:2017bld} $H$ and $H_2$ form a bi-doublet field (doublet under $SU(2)_L$ and $SU(2)_R$ of the third site). Although a bi-doublet can be formed by a doublet and its own dual, the presence of a different doublet $H_2$ allows the splitting between the top and bottom masses.

\paragraph{Potential.} Assuming a simplified 2HDM, the Higgs potential in Eq.~\eqref{eq:Higgs-potential} is augmented by
\begin{align}\label{eq:2HDMpotential}
V_H	& \supset  \mu_2^2 H_2^\dag H_2 
		+\frac{\lambda_2}{2} ( H_2^\dag H_2 )^2
		+ \lambda_3 ( H^\dag H ) ( H_2^\dag H_2 ) 
		+\lambda_4 ( H^\dag H_2 ) ( H_2^\dag H ) 
		+\frac{\lambda_5}{2} \left( ( H^\dag H_2 )^2 + \rm h.c \right)
\end{align}
and the mixing potential in Eq.~\eqref{eq:portal-potential} by
\begin{equation} \label{eq:portal-2HDMpotential}
    V_{\Omega H}  \supset 
    \eta_2\,  H_2^\dag H_2\,  \Omega_1^\dag \Omega_1
		+ \eta_4\,  H_2^\dag H_2\, \Tr [ \Omega_3^\dag \Omega_3 ]~.
\end{equation}
For the Yukawa couplings, the same terms with $H$ in Eq.~\eqref{eq:LYuk} can be written with $H_2$. 
\paragraph{Radial modes.}
In the Higgs basis where only $H$ takes a VEV, the two doublets can be parametrized as
\begin{equation} \label{eq:2HDM_decomposition}
    H= \begin{pmatrix}
            \eta^+_W \\
         \dfrac{v+h}{\sqrt2} + i\, \eta_Z
    \end{pmatrix}~, \qquad
    H_2= \begin{pmatrix}
         h^+\\
         \dfrac{1}{\sqrt 2} (h_R+i\, h_I) 
    \end{pmatrix}~,
\end{equation}
where $\eta_W^\pm, \eta_Z$ become the $W^\pm$ and $Z$ radial modes, respectively, and $h$ is the physical Higgs boson. The scalar radial degrees of freedom $h$, $h^\pm$, $h_R$ and $h_I$ aquire the following masses~\cite{Branco:2011iw,Belyaev:2016lok}
\begin{align} \label{eq:2HDM_spectrum}
\begin{aligned}
&m_h^2=\lambda v^2~,\\
&m_{\pm}^2=\mu_2^2+\frac{\lambda_3}{2} v^2~, \\
&m_R^2=\mu_2^2+\frac{\lambda_3 +\lambda_4+\lambda_5}{2}v^2~, \\
&m_I^2=\mu_2^2+\frac{\lambda_3 +\lambda_4-\lambda_5}{2}v^2~.
\end{aligned}
\end{align}
\paragraph{BFB conditions.}

For the potential \eqref{eq:2HDMpotential}, necessary and sufficient conditions to ensure boundedness from below were found to be~\cite{Klimenko:1984qx}
\begin{equation} \label{eq:2HDM_BFB}
    \lambda > 0~, \qquad \lambda_2 > 0~, \qquad \lambda_3 > -\sqrt{\lambda\, \lambda_2}~,\qquad \lambda_3+\lambda_4-|\lambda_5|>  -\sqrt{\lambda\, \lambda_2}~.
\end{equation}
For the mixed part of the potential $V_{\Omega H}$, similarly to Eq.~\eqref{eq:mix_BFB}, the derived conditions are
 \begin{align} 
    \begin{aligned} 
     & \eta_2 >- \sqrt{\lambda_2 \rho_1}~,\qquad 
     \eta_4 >- \min \left[\sqrt{\lambda_2 (\rho_2 + \rho_2')},\sqrt{\lambda_2 (3\rho_2 + \rho_2')} \right]~. 
    \end{aligned}
 \end{align}
 
\paragraph{$\beta$-functions.}
\label{app:2HDM_beta}

The $\beta$-functions above the 4321 breaking scale are presented in Appendix~\ref{app:4321_beta} with the gray terms to be included. Another small difference is that the coefficient in front of $y_t^3$ in Eq.~\eqref{eq:beta_yt} becomes $6$ instead of $11/2$.

Below 4321 breaking scale, we can run in the 2HDM until the mass threshold of the heavier Higgs. The most general Yukawa sector is 
\begin{align}
   - \mathcal{L}_{\rm Yuk}^{\rm (2HDM)} &= 
   y_{t}\, \bar q_L^{(3)} \tilde H u_R^{(3)} 
   + y_{b}\, \bar q_L^{(3)} H d_R^{(3)}
   + y_{\tau}\, \bar \ell_L^{(3)} H e_R^{(3)}
   \nonumber\\
   &+ y_{t2}\, \bar q_L^{(3)} \tilde H_2 u_R^{(3)} 
   + y_{b2}\, \bar q_L^{(3)} H_2 d_R^{(3)}
   + y_{\tau2}\, \bar \ell_L^{(3)} H_2 e_R^{(3)}~.
   \label{eq:Yukawa2HDM}
\end{align}
In the case $\langle H \rangle = v$ and $\langle H_2 \rangle =0$, we expect from UV matching $y_t = y_{b2}=y_{\tau2}$ and $y_{t2}=y_{b}=y_{\tau}$ (see Appendix~\ref{app:UV_Yukawa}).

The gauge and scalar contributions to the 2HDM $\beta$-functions are the same as in Eqs.(\ref{eq:betamH1}-\ref{eq:betamH2}) and (\ref{eq:betalambda1}-\ref{eq:betalambda5}) taking $\eta_{1,2,3,4} \rightarrow 0$.
The Yukawa contributions are changed as follow
\begin{align}
\begin{aligned} \label{eq:beta2HDMscalar}
    \beta_{\mu_H^2} &\supset 6 y_t^2 \mu_H^2~, 
     \qquad
    &\beta_{\mu_2^2} & \supset \gr{8 y_t^2 \mu_2^2 }~,
     \\
    \beta_{\lambda} &\supset 12 y_t^2 \lambda - 12 y_t^4~,
     \qquad
    &\beta_{\lambda_2} &\supset \gr{16 y_t^2 \lambda_2 - 16 y_t^4}~,
    \\
    \beta_{\lambda_3} &\supset 6 y_t^2 \lambda_3 \gr{\,+\, 8 y_t^2 \lambda_3 - 12 y_t^4}~,
      \qquad
    &
    \beta_{\lambda_4} &\supset 6 y_t^2 \lambda_4  \gr{\,+\, 8 y_t^2 \lambda_4 + 12 y_t^4}~,
     \\
    \beta_{\lambda_5} &\supset  6 y_t^2 \lambda_5 \gr{\,+\, 8 y_t^2 \lambda_5 }~.
    \end{aligned}
\end{align}
 where the $y_b$ contributions were neglected. 
The difference in multiplicative factors with respect to Eqs.~(\ref{eq:betamH1}-\ref{eq:betamH2}) and (\ref{eq:betalambda1}-\ref{eq:betalambda5}) is due to the fact that we neglected the neutrino Yukawa in equation Eq.~\eqref{eq:Yukawa2HDM}, whereas it is automatically included in the 4321 beta functions since both chiralities are part of $SU(4)$ multiplets.

The top Yukawa couplings in equation Eq.~\eqref{eq:Yukawa2HDM} run as
\begin{align}\label{eq:beta2HDMyukawa}
     \beta_{y_{t}} &= y_t\left( -\frac{17}{12} g_1^2 -\frac94 g_2^2 -8 g_3^2 +  \frac92 y_t^2 \gr{\,+\, \frac12 y_t^2} \right)~.
\end{align}
Without UV input, we could also consider a simpler 2HDM where only one of the Higgs couple to fermions, i.e. $y_{t2} = y_{b2}=y_{\tau2}=0$, in which case the gray terms in all equations above Eq.~\eqref{eq:beta2HDMscalar} and Eq.~\eqref{eq:beta2HDMyukawa} would be absent.

\subsection{Extended scalar sector}
\label{app:extended_scalars}

For our analysis of the UV behavior, we only considered $\Omega_1$ and $\Omega_3$. However, we mentioned that other scalars might be required for phenomenological reasons.
Below, we briefly review which additional scalars have been considered in the literature, their role and their contribution to the Lagrangian. 

\paragraph{$H_{2} \sim (\boldsymbol{1}, \boldsymbol 1, \boldsymbol 2)_{1/2}$} was already discussed in Appendix~\ref{app:2HDM}.

\paragraph{$H_{15} \sim (\boldsymbol{15}, \boldsymbol 1, \boldsymbol 2)_{1/2}$} from Refs.~\cite{Bordone:2018nbg,Cornella:2019hct} breaks the electroweak symmetry with a VEV aligned along $T^{15}\propto \text{diag}(1,1,1,-3)$. This field generates a splitting between the third generation quark and lepton in the Yukawa couplings allowing for $m_b \neq m_\tau$.\footnote{The $m_t\neq m_\nu$  splitting can in principle also be realized this way but would require an unnatural fine-tuning of $10^{-12}$. For a realistic neutrino mechanism see Ref.~\cite{Fuentes-Martin:2020pww}.}
Its Lagrangian extension is similar to the one of the second Higgs $H_2$. Three noticeable differences with the previous case are: 
\begin{enumerate}[i)]
    \item $H$ and $H_{15}$ cannot come from the same bi-doublet in the UV.
    \item Since it is an adjoint of $SU(4)$ it can only couple to the third generation fermions in the Yukawa, therefore exhibiting a $U(2)^5$ symmetry on the light generations. On the contrary $H$ and $H_2$ can couple to the light generations and their Yukawa couplings were just assumed to be small.
    \item It contains more fields in the broken phase of 4321 namely $H_{15} \rightarrow (\boldsymbol 8, \boldsymbol 2)_{1/2} \oplus (\boldsymbol 3, \boldsymbol 2)_{7/6} \oplus (\boldsymbol{\overline 3}, \boldsymbol 2)_{-1/6} \oplus (\boldsymbol 1, \boldsymbol 2)_{1/2}$. One can assume that the non-trivial $SU(3)_c$ representations decouple, retrieving the standard 2HDM again. Otherwise, the theory below 4321 breaking should also contain these modes.
\end{enumerate}

\paragraph{$\Omega_{15} \sim (\boldsymbol{15}, \boldsymbol 1, \boldsymbol 1)_0$} 
from Refs.~\cite{DiLuzio:2018zxy,Cornella:2019hct} breaks 4321 with the VEV $\langle \Omega_{15}\rangle = \omega_{15}\, T^{15}$.
Similarly to $H_{15}$, this field splits quarks and leptons.
This time, the splitting happens in the coupling of the third generation quarks and leptons to their vector-like partners, and after integrating out the vector-like fermions, to the $U_1$ leptoquark. The masses of the vector-like fermions are also split, allowing the vector-like leptons to be lighter than the vector-like quarks.
 The Yukawa terms can be written as
\begin{equation}
    -\mathcal{L} \supset \lambda_{15} \bar \Psi_L \Omega_{15} \chi_R + \lambda'_{15} \bar \chi_L \Omega_{15} \chi_R + \text{h.c.}
\end{equation}
This field could also be used to split the leptons and quarks masses directly, as with $H_{15}$, by introducing higher dimensional operators, e.g. $\frac{1}{\Lambda} \bar \Psi_L\, \Omega_{15} \,\tilde H \,\Psi_R^+$. In the broken phase of 4321, this field decomposes into $\Omega_{15}\to (\boldsymbol 1, \boldsymbol 1)_{0} \oplus (\boldsymbol 3, \boldsymbol 1)_{2/3} \oplus (\boldsymbol{
\overline 
3}, \boldsymbol 1)_{-2/3} \oplus (\boldsymbol 8, \boldsymbol 1)_{0}$.

Both $H_{15}$ and $\Omega_{15}$ have a more involved scalar potential which could give rise to the appearance of new Landau poles similarly to the $\Omega_{1,3}$ potential. 
Moreover their mixing coupling with the $\Omega_{1,3}$ sector will give contribution to the quartic couplings  studied in the main analysis with the same sign as the $g_4^4$ contribution in
Eqs.~(\ref{eq:betarho1}-\ref{eq:betarho4}). Their effect will therefore only strengthen the conclusions about the appearance of Landau poles.

Another important effect of extending the scalar content of the theory is its influence on the running of the gauge couplings as can be seen in Eq.~\eqref{eq:beta_couplings_formula}. At one loop, the condition for asymptotic freedom is $b_i>0$. The $\beta$-functions coefficients $b_i$ for $i=1,2,3,4$ with the minimal scalar content are presented in Eq.~\eqref{eq:beta_gauge_couplings}. Since the IR value of $g_4$ is large and drives most scalar quartic beta-functions, the faster it decreases in the UV, the later the Landau poles arise. Adding more scalars is equivalent to decreasing $b_4$ and slowing down its approach to asymptotic freedom. The influence of each scalar on the $b_i$ coefficients is presented in Table~\ref{tab:DynkinRep_scalars}. We can conclude that adding all these scalars to the theory is not enough to directly ruin the one-loop asymptotic freedom condition for gauge couplings, but they would accelerate the appearance of the Landau poles in the scalar sector as discussed in Section~\ref{sec:perturbativity}.

\begin{table}[H]
    \centering
    \begin{align*}
	\begin{array}{|c|c|c|c|c|c|c|c|}
	\hline
        S_2(R_s) & H & H_2 & H_{15} & \Omega_{15} & \Omega_3 & \Omega_1 \\
        \hline
        U(1) & 1 & 1 & 15 & 0 & 2/3 & 2 \\
        SU(2)_L & 1 & 1 & 15 & 0 & 0 & 0 \\
        SU(3) & 0 & 0 & 0 & 0 & 4 & 0 \\
        SU(4) & 0 & 0 & 16 & 8 & 3 & 1 \\
    \hline
    \end{array}
	\end{align*}
	\caption{Dynkin representation multiplicity factor of each scalar.}
	\label{tab:DynkinRep_scalars}
\end{table}

\section{Constraints from UV embedding}
\label{app:UV_embedding}
Assuming a PS$^3$ origin of the 4321 model as in~\cite{Bordone:2017bld} yields specific relations between the coefficient of the scalar potential and the Yukawa couplings. By considering the representations of the fields in this UV setup, their decompositions under the subgroup
\begin{equation}
    {\rm PS}_{1+2} \times {\rm PS}_3 \to SU(4)_3 \times SU(3)_{1+2} \times SU(2)_L \times U(1)'~,
\end{equation} 
where ${\rm PS}_i = SU(4)_i \times SU(2)_{L,i} \times SU(2)_{R,i}$, read as
\begin{align}
	\begin{aligned} \label{eq:UVdecomposition}
		\Phi \sim (1,2,\bar 2)_3 &\to H \sim (1,1,2)_{\frac12} + \tilde{H_2} \sim (1,1,2)_{-\frac12}~, \\
		\Omega \sim (4,2,1)_{1+2} \times (\bar 4, \bar 2, 1)_3  &\to  
		\Omega_3 \sim (\bar 4,3,1)_{\frac16} + \Omega_3' \sim (\bar 4,3,3)_{\frac16} \\
		&\quad  + \Omega_1 \sim (\bar 4,1,1)_{-\frac12} + \Omega_1' \sim (\bar 4,1,3)_{-\frac12}~,  \\
		\Psi_{R}^{(3)} \sim (4,1,2)_3 &\to \Psi_R^+ \sim (4,1,1)_\frac12 + \Psi_R^- \sim (4,1,1)_{-\frac12}~\,,\\
		\Psi_{L}^{(3)} \sim (4,2,1)_3 &\to \Psi_L \sim (4,2,1)_0~.
	\end{aligned}
\end{align}

\subsection{Yukawa sector} \label{app:UV_Yukawa}

We consider a Higgs bi-doublet field that couples mostly to the third generation of fermions. Neglecting light generations coupling, two Yukawas terms can be written with $\Phi^c =  \sigma_2 \Phi^*  \sigma_2$. 
The complex bi-doublet $\Phi$ can be written as
\begin{equation}
\Phi =\begin{pmatrix} \tilde H_2 & H \end{pmatrix} =\begin{pmatrix} {H_2^0}^* & H^+ \\ -H_2^- & H^0 \end{pmatrix}~, \qquad \Phi^c = \begin{pmatrix} \tilde H &  H_2  \end{pmatrix} =\begin{pmatrix} {H^0}^* & H_2^+ \\ -H^- & H_2^0 \end{pmatrix}~.
\end{equation}
where $ H $ and $ H_2 $ are $SU(2)_L$ doublets and $\tilde H_{(2)} = i\sigma^2 H_{(2)}^* $.
Since $\Psi_R^{(3)} = \begin{pmatrix}
	\Psi_R^+ \\ \Psi_R^-  \end{pmatrix}$ the Yukawa Lagrangian expands to
\begin{align}
	\begin{aligned}
		-\mathcal{L}_{\rm Yuk} &= \tilde y' \, \bar \Psi_L^{(3)} \Phi \Psi_R^{(3)} +  \tilde y \,\bar \Psi_L^{(3)} \Phi^c \Psi_R^{(3)} \\
		&= \tilde y'\, \bar \Psi_L \tilde H_2 \Psi_R^+ + \tilde y' \,  \bar \Psi_L H \Psi_R^- + \tilde y\, \bar \Psi_L \tilde H \Psi_R^+ + \tilde y \,  \bar \Psi_L H_2 \Psi_R^-~. 
	\end{aligned}
\end{align}
Generally, the Higgses will get the aligned VEVs
\begin{equation}
	\langle H \rangle =\begin{pmatrix}
		0 \\ \frac{v_1}{\sqrt 2}
	\end{pmatrix}~, \qquad
	\langle H_2 \rangle = \begin{pmatrix}
	    0 \\ \frac{v_2}{\sqrt 2}
	\end{pmatrix}~.
\end{equation}
and give rise to the masses
\begin{align}
	\begin{aligned}
		& m_t=m_\nu = \frac{1}{\sqrt 2}(\tilde y'\, v_2 + \tilde y \,v_1) = \frac{v}{\sqrt 2} \underbrace{(\tilde y'\, c_\alpha +\tilde y\, s_\alpha)}_{\displaystyle y_t}~, \\
		 &m_b = m_\tau = \frac{1}{\sqrt 2}(\tilde y'\, v_1 + \tilde y\, v_2)  = \frac{v}{\sqrt 2} \underbrace{(\tilde y'\, s_\alpha +\tilde y\, c_\alpha)}_{\displaystyle y_b}~.
	\end{aligned}
\end{align}
where we defined $v^2=v_1^2+v_2^2$ and $\tan \alpha = \dfrac{v_1}{v_2}$.
The choice of $\tan \alpha$ is arbitrary as one is free to choose the values of the Yukawa couplings $\tilde y$ and $\tilde y'$ in order to reproduce $m_t$ and $m_b$.
In the case where $v_1  = v, ~ v_2 = 0$, we have 
\begin{equation}\label{eq:UV_Yuk} 
    \tilde y= y_t \quad\text{and}\quad \tilde y'=y_b ~.
\end{equation}

\subsection{Scalar sector} \label{app:scalar_UV}

\paragraph{Higgs potential (2HDM).}

The most general potential for a bi-doublet Higgs is~\cite{Chauhan:2019fji}
\begin{align}
\begin{aligned}
	V_{\rm \Phi}^{\rm UV}= &~ - \tilde \mu_\Phi^2 \Tr[ \Phi^\dag \Phi ] - \left(\tilde \mu_{\Phi12}^2 \Tr[  \Phi^\dag \Phi^c] +\rm h.c.\right) \\
	&+ \tilde \lambda \Tr[ \Phi^\dag \Phi ]^2 +
	\left(\tilde \lambda_2 \Tr[\Phi^\dag \Phi^c  ]^2 + \rm h.c.\right) \\
	&+
	\tilde \lambda_3 \Tr[ \Phi^\dag \Phi^c  ]\Tr[ {\Phi^c}^\dag \Phi ] 
	+
	\left(\tilde \lambda_4 \Tr[ \Phi^\dag \Phi ] \Tr[ \Phi^\dag \Phi^c  ] + \rm h.c. \right)~.
\end{aligned}
\end{align}
Using the decomposition of the invariants
\begin{align}
	\begin{aligned}
	 \Tr[\Phi^\dag \Phi]&= H_2^\dag H_2 + H^\dag H~, \\
	 \Tr[\Phi^\dag \Phi^c ]&= \tilde H_2^\dag \tilde H + H^\dag H_2 = 2 H^\dag H_2~. 	
	\end{aligned}
\end{align}
the potential reduces to the well-known 2HDM 
\cite{Branco:2011iw} 
\begin{align}
	\begin{aligned}
		V_{\rm H}
		=& -\tilde \mu_\Phi^2 \left(H^\dag H + H_2^\dag H_2\right) - 2\left(\tilde \mu_{\Phi12}^2 H^\dag H_2 + \rm h.c.\right) \\ 
		&+ \tilde \lambda \left(  (H^\dag H)^2 + (H_2^\dag H_2)^2 + 2 (H^\dag H)(H_2^\dag H_2)  \right) \\
		&+ 4 \left( \tilde \lambda_2 (H^\dag H_2)^2  + \rm h.c. \right) 
		+ 4 \tilde \lambda_3 (H^\dag H_2)( H_2^\dag H)   \\
		&+ 2 \left(\tilde \lambda_4   (H_2^\dag H_2)( H_2^\dag H) + \tilde \lambda_4 (H^\dag H)( H^\dag H_2)  + \rm h.c.   \right)~. 
	\end{aligned}
\end{align}

Comparing with the simplified 2HDM potential Eq.~\eqref{eq:2HDMpotential}, we find $\tilde \mu_{\Phi12}= \tilde \lambda_4 = 0$ and  the relations between the coefficients coming from the UV unification is
\begin{equation}\label{eq:UV_mOmega}
	\mu_H=\mu_2
\end{equation}
and
\begin{equation} \label{eq:UV_lambda} 
	\lambda= \lambda_2 = \lambda_3 ~.
\end{equation}

\paragraph{$\Omega$ potential.}

The UV relations of the $\Omega$ potential coefficients would require considering all degrees of freedom from the decomposition in Eq.~\eqref{eq:UVdecomposition}. The next step would be to decouple triplets of $SU(2)_L$, so that $\Omega$ only takes a VEV along the $SU(2)_L$-preserving direction. Neglecting the contribution from integrating out $\Omega_{1,3}'$, we write a multiplet where these degrees of freedom are omitted from the start, as well as the $SU(2)_{\ell L}$ and $SU(2)_{3 L}$ indices,
\begin{equation}
	\Omega=  \begin{pmatrix}
		\Omega_3 \\ \Omega_1
	\end{pmatrix} \mathbb{1}_{2L} + \begin{pmatrix}
		\Omega'_3 \\ \Omega'_1
	\end{pmatrix}_I \tau^I_{2L}
	\approx \begin{pmatrix}
		\Omega_3 \\ \Omega_1
	\end{pmatrix}~,
\end{equation}
where $\Omega_3$ and $\Omega_1$ are both anti-fundamental of $SU(4)_3$ and the vector is an $SU(4)_\ell$ multiplet.
Keeping only the two $SU(4)$ indices, the different invariants composing the potential read
\begin{align}
	\begin{aligned}
V_{\Omega}^{\rm UV} =~& -\tilde \mu_\Omega \Tr [ \Omega^\dag \Omega] 
	 + \frac{\tilde \rho_1}{2} \Tr [ \Omega^\dag \Omega]^2 
	 +  \frac{\tilde \rho_2}{2} \Tr[ \Omega^\dag \Omega \Omega^\dag \Omega]~,
	\\
	&+ \frac{\tilde \rho_3}{4!} \, \epsilon_{\alpha \beta\gamma\sigma}\epsilon^{abcd} (\Omega)^\alpha_a (\Omega)^\beta_b (\Omega)^\gamma_c (\Omega)^\sigma_d~,
	\end{aligned}
\end{align}
which yields
\begin{align}
	\begin{aligned}
		V_{\Omega}
		=& -\tilde \mu_\Omega \left(\Tr[\Omega_3^\dag \Omega_3] + \Omega_1^\dag \Omega_1\right) \\
		& +\frac{\tilde \rho_1}{2} \left(\Tr[\Omega_3^\dag \Omega_3]^2 + 2\Tr[\Omega_3^\dag \Omega_3] \Omega_1^\dag \Omega_1+(\Omega_1^\dag \Omega_1)^2\right) \\ 
		& +\frac{\tilde \rho_2}{2} \left(\Tr[\Omega_3^\dag \Omega_3\Omega_3^\dag \Omega_3] + 2  \Omega_1^\dag  \Omega_3 \Omega_3^\dag  \Omega_1+(\Omega_1^\dag \Omega_1)^2\right) \\
		& + \frac{\tilde \rho_3}{3!} \, \epsilon_{\alpha \beta\gamma\sigma} \epsilon^{abc}(\Omega_3)^\alpha_a (\Omega_3)^\beta_b (\Omega_3)^\gamma_c (\Omega_1)^\sigma~.
	\end{aligned}
\end{align}
The coefficients in the potential Eq.~\eqref{eq:Omegapotential} are connected because of their UV origin through the relations
\begin{equation}\label{eq:UV_rho} 
	\rho_1=\rho_2 + \rho_2'~, \qquad \rho_2=\rho_3~, \qquad \rho_2'=\rho_4 
\end{equation}
and 
\begin{equation}\label{eq:UV_mH} 
	m_{\Omega_1}=m_{\Omega_3}~.
\end{equation}
    
\paragraph{Mixed $\Omega$-Higgses potential.}

As for the mixed potential Eqs.~\eqref{eq:portal-potential} and \eqref{eq:portal-2HDMpotential}, noticing that all mixed term come from a single UV term (where again the $SU(2)_L$-triplet contributions were neglected):
\begin{equation}
	V_{\Omega \Phi}^{\rm UV} = \tilde \eta ~ \Tr [\Omega^\dag \Omega] \Tr[\Phi^\dag \Phi]~,
\end{equation}
we find the UV relation between the coefficients in $V_{\Omega H}$
\begin{equation} \label{eq:UV_eta} 
	\eta_1=\eta_2= \eta_3 = \eta_4~.
\end{equation}

\newpage

\bibliographystyle{ieeetr}
\bibliography{4321RGE.bib}

\begin{thebibliography}{10}

\bibitem{LHCb:2014vgu}
R.~Aaij {\em et~al.}, ``{Test of lepton universality using $B^{+}\rightarrow
  K^{+}\ell^{+}\ell^{-}$ decays},'' {\em Phys. Rev. Lett.}, vol.~113,
  p.~151601, 2014.

\bibitem{LHCb:2017avl}
R.~Aaij {\em et~al.}, ``{Test of lepton universality with $B^{0} \rightarrow
  K^{*0}\ell^{+}\ell^{-}$ decays},'' {\em JHEP}, vol.~08, p.~055, 2017.

\bibitem{LHCb:2019hip}
R.~Aaij {\em et~al.}, ``{Search for lepton-universality violation in $B^+\to
  K^+\ell^+\ell^-$ decays},'' {\em Phys. Rev. Lett.}, vol.~122, no.~19,
  p.~191801, 2019.

\bibitem{LHCb:2021trn}
R.~Aaij {\em et~al.}, ``{Test of lepton universality in beauty-quark decays},''
  {\em Nature Phys.}, vol.~18, no.~3, pp.~277--282, 2022.

\bibitem{BaBar:2012obs}
J.~P. Lees {\em et~al.}, ``{Evidence for an excess of $\bar{B} \to D^{(*)}
  \tau^-\bar{\nu}_\tau$ decays},'' {\em Phys. Rev. Lett.}, vol.~109, p.~101802,
  2012.

\bibitem{BaBar:2013mob}
J.~P. Lees {\em et~al.}, ``{Measurement of an Excess of $\bar{B} \to
  D^{(*)}\tau^- \bar{\nu}_\tau$ Decays and Implications for Charged Higgs
  Bosons},'' {\em Phys. Rev. D}, vol.~88, no.~7, p.~072012, 2013.

\bibitem{Belle:2015qfa}
M.~Huschle {\em et~al.}, ``{Measurement of the branching ratio of $\bar{B} \to
  D^{(\ast)} \tau^- \bar{\nu}_\tau$ relative to $\bar{B} \to D^{(\ast)} \ell^-
  \bar{\nu}_\ell$ decays with hadronic tagging at Belle},'' {\em Phys. Rev. D},
  vol.~92, no.~7, p.~072014, 2015.

\bibitem{LHCb:2015gmp}
R.~Aaij {\em et~al.}, ``{Measurement of the ratio of branching fractions
  $\mathcal{B}(\bar{B}^0 \to
  D^{*+}\tau^{-}\bar{\nu}_{\tau})/\mathcal{B}(\bar{B}^0 \to
  D^{*+}\mu^{-}\bar{\nu}_{\mu})$},'' {\em Phys. Rev. Lett.}, vol.~115, no.~11,
  p.~111803, 2015.
\newblock [Erratum: Phys.Rev.Lett. 115, 159901 (2015)].

\bibitem{LHCb:2017smo}
R.~Aaij {\em et~al.}, ``{Measurement of the ratio of the $B^0 \to D^{*-} \tau^+
  \nu_{\tau}$ and $B^0 \to D^{*-} \mu^+ \nu_{\mu}$ branching fractions using
  three-prong $\tau$-lepton decays},'' {\em Phys. Rev. Lett.}, vol.~120,
  no.~17, p.~171802, 2018.

\bibitem{LHCb:2017rln}
R.~Aaij {\em et~al.}, ``{Test of Lepton Flavor Universality by the measurement
  of the $B^0 \to D^{*-} \tau^+ \nu_{\tau}$ branching fraction using
  three-prong $\tau$ decays},'' {\em Phys. Rev. D}, vol.~97, no.~7, p.~072013,
  2018.

\bibitem{Alonso:2015sja}
R.~Alonso, B.~Grinstein, and J.~Martin~Camalich, ``{Lepton universality
  violation and lepton flavor conservation in $B$-meson decays},'' {\em JHEP},
  vol.~10, p.~184, 2015.

\bibitem{Calibbi:2015kma}
L.~Calibbi, A.~Crivellin, and T.~Ota, ``{Effective Field Theory Approach to
  $b\to s\ell\ell^{(')}$, $B\to K^{(*)}\nu\overline{\nu}$ and $B\to
  D^{(*)}\tau\nu$ with Third Generation Couplings},'' {\em Phys. Rev. Lett.},
  vol.~115, p.~181801, 2015.

\bibitem{Barbieri:2015yvd}
R.~Barbieri, G.~Isidori, A.~Pattori, and F.~Senia, ``{Anomalies in $B$-decays
  and $U(2)$ flavour symmetry},'' {\em Eur. Phys. J. C}, vol.~76, no.~2, p.~67,
  2016.

\bibitem{Buttazzo:2017ixm}
D.~Buttazzo, A.~Greljo, G.~Isidori, and D.~Marzocca, ``{B-physics anomalies: a
  guide to combined explanations},'' {\em JHEP}, vol.~11, p.~044, 2017.

\bibitem{Angelescu:2021lln}
A.~Angelescu, D.~Be\v{c}irevi\'c, D.~A. Faroughy, F.~Jaffredo, and
  O.~Sumensari, ``{Single leptoquark solutions to the B-physics anomalies},''
  {\em Phys. Rev. D}, vol.~104, no.~5, p.~055017, 2021.

\bibitem{Pati:1974yy}
J.~C. Pati and A.~Salam, ``{Lepton Number as the Fourth Color},'' {\em Phys.
  Rev. D}, vol.~10, pp.~275--289, 1974.
\newblock [Erratum: Phys.Rev.D 11, 703--703 (1975)].

\bibitem{Diaz:2017lit}
B.~Diaz, M.~Schmaltz, and Y.-M. Zhong, ``{The leptoquark
  Hunter\textquoteright{}s guide: Pair production},'' {\em JHEP}, vol.~10,
  p.~097, 2017.

\bibitem{DiLuzio:2017vat}
L.~Di~Luzio, A.~Greljo, and M.~Nardecchia, ``{Gauge leptoquark as the origin of
  B-physics anomalies},'' {\em Phys. Rev. D}, vol.~96, no.~11, p.~115011, 2017.

\bibitem{Barbieri:2017tuq}
R.~Barbieri and A.~Tesi, ``{$B$-decay anomalies in Pati-Salam SU(4)},'' {\em
  Eur. Phys. J. C}, vol.~78, no.~3, p.~193, 2018.

\bibitem{Calibbi:2017qbu}
L.~Calibbi, A.~Crivellin, and T.~Li, ``{Model of vector leptoquarks in view of
  the $B$-physics anomalies},'' {\em Phys. Rev. D}, vol.~98, no.~11, p.~115002,
  2018.

\bibitem{Fornal:2018dqn}
B.~Fornal, S.~A. Gadam, and B.~Grinstein, ``{Left-Right SU(4) Vector Leptoquark
  Model for Flavor Anomalies},'' {\em Phys. Rev. D}, vol.~99, no.~5, p.~055025,
  2019.

\bibitem{Blanke:2018sro}
M.~Blanke and A.~Crivellin, ``{$B$ Meson Anomalies in a Pati-Salam Model within
  the Randall-Sundrum Background},'' {\em Phys. Rev. Lett.}, vol.~121, no.~1,
  p.~011801, 2018.

\bibitem{DiLuzio:2018zxy}
L.~Di~Luzio, J.~Fuentes-Martin, A.~Greljo, M.~Nardecchia, and S.~Renner,
  ``{Maximal Flavour Violation: a Cabibbo mechanism for leptoquarks},'' {\em
  JHEP}, vol.~11, p.~081, 2018.

\bibitem{Greljo:2018tuh}
A.~Greljo and B.~A. Stefanek, ``{Third family quark\textendash{}lepton
  unification at the TeV scale},'' {\em Phys. Lett. B}, vol.~782, pp.~131--138,
  2018.

\bibitem{Bordone:2018nbg}
M.~Bordone, C.~Cornella, J.~Fuentes-Mart\'\i{}n, and G.~Isidori, ``{Low-energy
  signatures of the $\mathrm{PS}^3$ model: from $B$-physics anomalies to
  LFV},'' {\em JHEP}, vol.~10, p.~148, 2018.

\bibitem{Cornella:2019hct}
C.~Cornella, J.~Fuentes-Martin, and G.~Isidori, ``{Revisiting the vector
  leptoquark explanation of the B-physics anomalies},'' {\em JHEP}, vol.~07,
  p.~168, 2019.

\bibitem{Fuentes-Martin:2020bnh}
J.~Fuentes-Mart\'\i{}n and P.~Stangl, ``{Third-family quark-lepton unification
  with a fundamental composite Higgs},'' {\em Phys. Lett. B}, vol.~811,
  p.~135953, 2020.

\bibitem{Guadagnoli:2020tlx}
D.~Guadagnoli, M.~Reboud, and P.~Stangl, ``{The Dark Side of 4321},'' {\em
  JHEP}, vol.~10, p.~084, 2020.

\bibitem{Fuentes-Martin:2020hvc}
J.~Fuentes-Mart\'\i{}n, G.~Isidori, M.~K\"onig, and N.~Selimovi\'c, ``{Vector
  Leptoquarks Beyond Tree Level III: Vector-like Fermions and Flavor-Changing
  Transitions},'' {\em Phys. Rev. D}, vol.~102, p.~115015, 2020.

\bibitem{Baker:2021llj}
M.~J. Baker, D.~A. Faroughy, and S.~Trifinopoulos, ``{Collider signatures of
  coannihilating dark matter in light of the B-physics anomalies},'' {\em
  JHEP}, vol.~11, p.~084, 2021.

\bibitem{Bordone:2017bld}
M.~Bordone, C.~Cornella, J.~Fuentes-Martin, and G.~Isidori, ``{A three-site
  gauge model for flavor hierarchies and flavor anomalies},'' {\em Phys. Lett.
  B}, vol.~779, pp.~317--323, 2018.

\bibitem{Fuentes-Martin:2020pww}
J.~Fuentes-Martin, G.~Isidori, J.~Pag\`es, and B.~A. Stefanek, ``{Flavor
  non-universal Pati-Salam unification and neutrino masses},'' {\em Phys. Lett.
  B}, vol.~820, p.~136484, 2021.

\bibitem{Fuentes-Martin:2022xnb}
J.~Fuentes-Martin, G.~Isidori, J.~M. Lizana, N.~Selimovic, and B.~A. Stefanek,
  ``{Flavor hierarchies, flavor anomalies, and Higgs mass from a warped extra
  dimension},'' 3 2022.

\bibitem{King:2021jeo}
S.~F. King, ``{Twin Pati-Salam theory of flavour with a TeV scale vector
  leptoquark},'' {\em JHEP}, vol.~11, p.~161, 2021.

\bibitem{Ibanez:1982fr}
L.~E. Ibanez and G.~G. Ross, ``{SU(2)-L x U(1) Symmetry Breaking as a Radiative
  Effect of Supersymmetry Breaking in Guts},'' {\em Phys. Lett. B}, vol.~110,
  pp.~215--220, 1982.

\bibitem{Inoue:1982pi}
K.~Inoue, A.~Kakuto, H.~Komatsu, and S.~Takeshita, ``{Aspects of Grand Unified
  Models with Softly Broken Supersymmetry},'' {\em Prog. Theor. Phys.},
  vol.~68, p.~927, 1982.
\newblock [Erratum: Prog.Theor.Phys. 70, 330 (1983)].

\bibitem{Ibanez:1982ee}
L.~E. Ibanez, ``{Locally Supersymmetric SU(5) Grand Unification},'' {\em Phys.
  Lett. B}, vol.~118, pp.~73--78, 1982.

\bibitem{Ellis:1982wr}
J.~R. Ellis, D.~V. Nanopoulos, and K.~Tamvakis, ``{Grand Unification in Simple
  Supergravity},'' {\em Phys. Lett. B}, vol.~121, pp.~123--129, 1983.

\bibitem{Ellis:1983bp}
J.~R. Ellis, J.~S. Hagelin, D.~V. Nanopoulos, and K.~Tamvakis, ``{Weak Symmetry
  Breaking by Radiative Corrections in Broken Supergravity},'' {\em Phys. Lett.
  B}, vol.~125, p.~275, 1983.

\bibitem{Alvarez-Gaume:1983drc}
L.~Alvarez-Gaume, J.~Polchinski, and M.~B. Wise, ``{Minimal Low-Energy
  Supergravity},'' {\em Nucl. Phys. B}, vol.~221, p.~495, 1983.

\bibitem{Babu:2016gpg}
K.~S. Babu, I.~Gogoladze, and S.~Khan, ``{Radiative Electroweak Symmetry
  Breaking in Standard Model Extensions},'' {\em Phys. Rev. D}, vol.~95, no.~9,
  p.~095013, 2017.

\bibitem{Coleman:1973jx}
S.~R. Coleman and E.~J. Weinberg, ``{Radiative Corrections as the Origin of
  Spontaneous Symmetry Breaking},'' {\em Phys. Rev. D}, vol.~7, pp.~1888--1910,
  1973.

\bibitem{Fuentes-Martin:2020luw}
J.~Fuentes-Mart\'\i{}n, G.~Isidori, M.~K\"onig, and N.~Selimovi\'c, ``{Vector
  leptoquarks beyond tree level. II. $\mathcal{O}(\alpha_s)$ corrections and
  radial modes},'' {\em Phys. Rev. D}, vol.~102, no.~3, p.~035021, 2020.

\bibitem{Elias-Miro:2012eoi}
J.~Elias-Miro, J.~R. Espinosa, G.~F. Giudice, H.~M. Lee, and A.~Strumia,
  ``{Stabilization of the Electroweak Vacuum by a Scalar Threshold Effect},''
  {\em JHEP}, vol.~06, p.~031, 2012.

\bibitem{Kannike:2012pe}
K.~Kannike, ``{Vacuum Stability Conditions From Copositivity Criteria},'' {\em
  Eur. Phys. J. C}, vol.~72, p.~2093, 2012.

\bibitem{Kannike:2016fmd}
K.~Kannike, ``{Vacuum Stability of a General Scalar Potential of a Few
  Fields},'' {\em Eur. Phys. J. C}, vol.~76, no.~6, p.~324, 2016.
\newblock [Erratum: Eur.Phys.J.C 78, 355 (2018)].

\bibitem{Chauhan:2019fji}
G.~Chauhan, ``{Vacuum Stability and Symmetry Breaking in Left-Right Symmetric
  Model},'' {\em JHEP}, vol.~12, p.~137, 2019.

\bibitem{ParticleDataGroup:2020ssz}
P.~A. Zyla {\em et~al.}, ``{Review of Particle Physics},'' {\em PTEP},
  vol.~2020, no.~8, p.~083C01, 2020.

\bibitem{Cornella:2021sby}
C.~Cornella, D.~A. Faroughy, J.~Fuentes-Martin, G.~Isidori, and M.~Neubert,
  ``{Reading the footprints of the B-meson flavor anomalies},'' {\em JHEP},
  vol.~08, p.~050, 2021.

\bibitem{Baker:2019sli}
M.~J. Baker, J.~Fuentes-Mart\'\i{}n, G.~Isidori, and M.~K\"onig, ``{High- $p_T$
  signatures in vector\textendash{}leptoquark models},'' {\em Eur. Phys. J. C},
  vol.~79, no.~4, p.~334, 2019.

\bibitem{Bai:2018jsr}
Y.~Bai and B.~A. Dobrescu, ``{Collider Tests of the Renormalizable Coloron
  Model},'' {\em JHEP}, vol.~04, p.~114, 2018.

\bibitem{Ilnicka:2018def}
A.~Ilnicka, T.~Robens, and T.~Stefaniak, ``{Constraining Extended Scalar
  Sectors at the LHC and beyond},'' {\em Mod. Phys. Lett. A}, vol.~33,
  no.~10n11, p.~1830007, 2018.

\bibitem{Fuentes-Martin:2019ign}
J.~Fuentes-Mart\'\i{}n, G.~Isidori, M.~K\"onig, and N.~Selimovi\'c, ``{Vector
  Leptoquarks Beyond Tree Level},'' {\em Phys. Rev. D}, vol.~101, no.~3,
  p.~035024, 2020.

\bibitem{Barbieri:1987fn}
R.~Barbieri and G.~F. Giudice, ``{Upper Bounds on Supersymmetric Particle
  Masses},'' {\em Nucl. Phys. B}, vol.~306, pp.~63--76, 1988.

\bibitem{Allwicher:2020esa}
L.~Allwicher, G.~Isidori, and A.~E. Thomsen, ``{Stability of the Higgs Sector
  in a Flavor-Inspired Multi-Scale Model},'' {\em JHEP}, vol.~01, p.~191, 2021.

\bibitem{Haller:2018nnx}
J.~Haller, A.~Hoecker, R.~Kogler, K.~M\"onig, T.~Peiffer, and J.~Stelzer,
  ``{Update of the global electroweak fit and constraints on two-Higgs-doublet
  models},'' {\em Eur. Phys. J. C}, vol.~78, no.~8, p.~675, 2018.

\bibitem{Azatov:2018kzb}
A.~Azatov, D.~Barducci, D.~Ghosh, D.~Marzocca, and L.~Ubaldi, ``{Combined
  explanations of B-physics anomalies: the sterile neutrino solution},'' {\em
  JHEP}, vol.~10, p.~092, 2018.

\bibitem{Marzocca:2021azj}
D.~Marzocca and S.~Trifinopoulos, ``{Minimal Explanation of Flavor Anomalies:
  B-Meson Decays, Muon Magnetic Moment, and the Cabibbo Angle},'' {\em Phys.
  Rev. Lett.}, vol.~127, no.~6, p.~061803, 2021.

\bibitem{Barbieri:2016las}
R.~Barbieri, C.~W. Murphy, and F.~Senia, ``{B-decay Anomalies in a Composite
  Leptoquark Model},'' {\em Eur. Phys. J. C}, vol.~77, no.~1, p.~8, 2017.

\bibitem{Marzocca:2018wcf}
D.~Marzocca, ``{Addressing the B-physics anomalies in a fundamental Composite
  Higgs Model},'' {\em JHEP}, vol.~07, p.~121, 2018.

\bibitem{Thomsen:2021ncy}
A.~E. Thomsen, ``{Introducing RGBeta: a Mathematica package for the evaluation
  of renormalization group $ \beta $-functions},'' {\em Eur. Phys. J. C},
  vol.~81, no.~5, p.~408, 2021.

\bibitem{Branco:2011iw}
G.~C. Branco, P.~M. Ferreira, L.~Lavoura, M.~N. Rebelo, M.~Sher, and J.~P.
  Silva, ``{Theory and phenomenology of two-Higgs-doublet models},'' {\em Phys.
  Rept.}, vol.~516, pp.~1--102, 2012.

\bibitem{Belyaev:2016lok}
A.~Belyaev, G.~Cacciapaglia, I.~P. Ivanov, F.~Rojas-Abatte, and M.~Thomas,
  ``{Anatomy of the Inert Two Higgs Doublet Model in the light of the LHC and
  non-LHC Dark Matter Searches},'' {\em Phys. Rev. D}, vol.~97, no.~3,
  p.~035011, 2018.

\bibitem{Klimenko:1984qx}
K.~G. Klimenko, ``{On Necessary and Sufficient Conditions for Some Higgs
  Potentials to Be Bounded From Below},'' {\em Theor. Math. Phys.}, vol.~62,
  pp.~58--65, 1985.

\end{thebibliography}

\end{document}